\documentclass[aps,pre,twocolumn,showpacs]{revtex4-1}

\usepackage{amssymb}
\usepackage{amsmath}
\usepackage{amsthm}
\usepackage{graphicx}
\usepackage{amsfonts}
\usepackage{color}
\usepackage{times}
\usepackage{natbib}
\usepackage{soul}
\usepackage{booktabs}
\usepackage{pifont}

\usepackage{hyperref}
\hypersetup{colorlinks=true}
\usepackage{mathpazo}

\RequirePackage{fix-cm}

\newcommand{\myskip}[1]{}   
\newcommand{\BEQ}{\begin{eqnarray}}      
\newcommand{\EEQ}{\end{eqnarray}}      
\newcommand{\BEA}{\begin{eqnarray}}      
\newcommand{\EEA}{\end{eqnarray}}      
\newcommand{\nn}{\nonumber }      
      
\newcommand{\half}{\frac{1}{2}}

\newcommand{\tr}{{\rm tr}}
\newcommand{\Tr}{{\rm Tr}}

\newcommand{\ruu}{r_{\uparrow \uparrow}}
\newcommand{\rud}{r_{\uparrow \downarrow}}

\newcommand{\rdd}{r_{\downarrow \downarrow}}
\newcommand{\Puu}{P_{\uparrow \uparrow}}
\newcommand{\Pud}{P_{\uparrow \downarrow}}
\newcommand{\Pdu}{P_{\downarrow \uparrow}}
\newcommand{\Pdd}{P_{\downarrow \downarrow}}

\newcommand{\ketup}{|\hspace{-1mm}\uparrow\rangle}
\newcommand{\ketdown}{|\hspace{-1mm}\downarrow\rangle}
\newcommand{\braup}{\langle\uparrow\hspace{-1mm}|}
\newcommand{\bradown}{\langle\downarrow\hspace{-1mm}|}
\newcommand{\keti}{|i \rangle}
\newcommand{\ketj}{|j \rangle}
\newcommand{\brai}{\langle i|}
\newcommand{\braj}{\langle j |}
\newcommand{\ketupu}{|\hspace{-1mm}\uparrow_{\bf u}\rangle}
\newcommand{\ketdownu}{|\hspace{-1mm}\downarrow_{\bf u}\rangle}
\newcommand{\braupu}{\langle\uparrow_{\bf u}\hspace{-1mm}|}
\newcommand{\bradownu}{\langle\downarrow_{\bf u}\hspace{-1mm}|}
\newcommand{\ketiu}{|i_{\bf u} \rangle}
\newcommand{\ketju}{|j_{\bf u} \rangle}
\newcommand{\braiu}{\langle i_{\bf u}|}
\newcommand{\braju}{\langle j_{\bf u} |}

\newcommand{\ketupee}{|\hspace{-1.1mm}\uparrow_{\epsilon \epsilon'}\rangle}

\newcommand{\braupee}{\langle\uparrow_{\epsilon \epsilon'}\hspace{-1mm}|}

\newcommand{\ketupepp}{|\hspace{-1mm}\uparrow_{\text{++}}\rangle}

\newcommand{\braupepp}{\langle\uparrow_{\text{++}}\hspace{-1mm}|}

\newcommand{\ketupemp}{|\hspace{-1mm}\uparrow_{\text{-\hspace{0.2mm}+}}\rangle}

\newcommand{\braupemp}{\langle\uparrow_{\text{-\hspace{0.2mm}+}}\hspace{-1mm}|}

\newcommand{\ketupepm}{|\hspace{-1mm}\uparrow_{\text{+\hspace{0.2mm}-}}\rangle}

\newcommand{\braupepm}{\langle\uparrow_{\text{+\hspace{0.2mm}-}}\hspace{-1mm}|}

\newcommand{\ketupemm}{|\hspace{-1mm}\uparrow_{\text{-\hspace{0.2mm}-}}\rangle}

\newcommand{\braupemm}{\langle\uparrow_{\text{-\hspace{0.2mm}-}}\hspace{-1mm}|}

\newcommand{\Ppp}{p_{\text{++}}}
\newcommand{\Ppm}{p_{\text{+\hspace{0.2mm}-}}}
\newcommand{\Pmp}{p_{\text{-\hspace{0.2mm}+}}}
\newcommand{\Pmm}{p_{\text{-\hspace{0.2mm}-}}}
\newcommand{\Pee}{p_{\epsilon \epsilon'}}

\newcommand{\Idd}{\mathbb{I}}

\usepackage{graphicx}
\usepackage{dcolumn}
\usepackage{bm}
\usepackage{mathtools}
\usepackage{lipsum}

\renewcommand{\thesection}{\arabic{section}}
\renewcommand{\theequation}{\thesection.\arabic{equation}}
\renewcommand{\thesection}{\arabic{section}}
\renewcommand{\thesubsection}{\thesection\arabic{subsection}}
\renewcommand{\theequation}{\thesection\arabic{equation}}

\begin{document}

\title{
Simultaneous measurement of two non-commuting quantum variables: Solution of a dynamical model}

\author{Mart{\'i} Perarnau-Llobet}
\email{marti.perarnau@rzg.mpg.de} 
\affiliation{Max-Planck-Institut f\"ur Quantenoptik, Hans-Kopfermann-Str. 1, D-85748 Garching, Germany}
\affiliation{ICFO-Institut de Ciencies Fotoniques, The Barcelona Institute of Science and Technology, 08860 Castelldefels, Barcelona, Spain}


\author{Theodorus Maria Nieuwenhuizen}
\affiliation{Institute for Theoretical Physics, University of Amsterdam, Science Park 904, 1090 GL  Amsterdam, The Netherlands} 
\affiliation{International Institute of Physics, UFRG,  Anel Vi\'ario da UFRN - Lagoa Nova, Natal - RN, 59064-741, Brazil}




\begin{abstract}
The possibility of performing simultaneous measurements in quantum mechanics is investigated in the context of the Curie-Weiss model for a projective measurement. 
 Concretely, we consider a spin--$\half$ system simultaneously interacting with two magnets, which act as measuring apparatuses of two different spin components. We work out the dynamics of this process and determine the final state of the measuring apparatuses, from which we can find  the probabilities of the four possible  outcomes of the measurements.  The measurement is found to be non-ideal, as (i) the joint statistics do not coincide with the one obtained by separately measuring each spin component, and (ii) the density matrix of 
the spin does not collapse in either of the measured observables. However, we give an operational interpretation of the process as a generalized quantum measurement, and show that it is  fully informative: The expected value of the measured spin components can be found with arbitrary precision for sufficiently many runs of the experiment.
\end{abstract}
 
 

\maketitle

 \setcounter{section}{0}
 \renewcommand{\thesection}{\arabic{section}}
 \section*{ Introduction}
 \setcounter{equation}{0} \setcounter{figure}{0}
 \renewcommand{\thesection}{\arabic{section}.}


Projective quantum measurements are usually described as an  instantaneous evolution, where the wavefunction \emph{collapses}  to an eigenstate of the measured observable. Yet progress in the last decades  have shown how physical mechanisms, such as decoherence and dephasing, might be responsible for this apparent collapse. In this case, the measurement postulate appears as a consequence of the particular interaction between system and apparatus, as well as the macroscopic size of the latter   (see \cite{OpusABN,OpusII,other1,other2,other3} and references therein). 
By treating the measurement as a physical evolution, in this work we explore the possibility of measuring simultaneously two non-commuting observables.
 We note that, while simultaneous measurements are usually not covered by the standard postulates of quantum mechanics, they are attempted experimentally  (see, e.g., the recent experiments \cite{Shay,gills}), and are a subject of high interest for the foundations of quantum physics \cite{OpusABN,Arthurs,Buschh,BuschLahti,MuynkI,MuynkII,Armen,Ziman,Ozawa,Andersson,Mohseni,ArmenII,Busch,BuschReview,Branciard,Bell,Wolf,Marco,Uola}.


In this article, we study joint measurements of two spin components, in which each measurement when treated individually corresponds to a projective measurement. While the statistics of joint qubit measurements is by now well understood (see \cite{BuschQubit,BranciardQubit} and references therein), here we focus our attention in the dynamics of such measurements, which allows us to explicitly show the  disturbance that the two apparatuses induce to each other and on  the system.  

 In order to describe the dynamics of the measurement, we use the  Curie Weiss model \cite{ei,eii,eiii}, which can be used to describe a projective quantum measurement of a spin--$\half$ system by a magnet \cite{OpusABN,ABNCW}. In this model, the magnet, which is in contact with a thermal bath, is initially set in a metastable paramagnetic state. The measurement then takes place when the interaction with the spin-$\half$ system triggers the magnet towards one of its two stable ferromagnetic states. These two robust, stable ferromagnetic states are identified with the pointer states of the apparatus. Following the initial attempts in \cite{OpusABN,Armen},  we study in detail the evolution of a spin-$\half$ system simultaneously interacting with two such magnets. 

We observe a competition between the two apparatuses, each of them trying to obtain information about a different component. This results into a non-ideal measurement:  The marginal probability distribution for the outcomes obtained by each apparatus does not correspond to the one given by the Born rule, and the spin does not collapse in either of the measured observables. Yet the joint measurement can be well described as a generalized quantum measurement, defined by a Positive-Operator-Valued-Measure (POVM). We also show that the expectation value of each spin component of the tested spin can be inferred after many runs of the process. 

It is important to stress that the whole measurement process, from the collapse of the wave function to the amplification of the microscopic signal, is here treated explicitly as a physical evolution between the tested system and the two measuring apparatuses. This allows us to describe how the system and both apparatuses are progressively disturbed by each other, leading to many features of non-ideal measurements. In this way, we complement previous studies on simultaneous measurements, which range from theoretic considerations on the possible statistics \cite{Ziman,Ozawa,Andersson,Busch,BuschReview,Branciard,Bell,Wolf,Marco,Uola}, to studies of specific measurement models  (see \cite{BuschLahti,Buschh} and references therein), including  the Von Neumann measurement set-up \cite{Arthurs}, continuous measurements \cite{Ruskov,Jordan,Shay,Luis} and weak measurements \cite{Mitchison,gills}.

The paper is structured as follows. In Sec. \ref{SecCW} we present the Curie-Weiss model for a quantum projective measurement. In Sec. \ref{Sec2app}, we explore the possibility of performing a simultaneous measurement. In the main text, we present a qualitative analysis based on free energy functions, which allows us to infer the final form of the apparatuses after the measurement, which is complemented by a detailed calculation in the Appendices  of the equations of motion of the process. Finally, in Sec. \ref{SecPOVM}  we provide an operational interpretation of the measurement using the theory of generalized quantum measurements. 
The paper closes with a discussion. Technical details are deferred to the Appendix.

 \setcounter{section}{0}
 \renewcommand{\thesection}{\arabic{section}}
\section{The Curie-Weiss model for a quantum process} 
 \setcounter{equation}{0} \setcounter{figure}{0}
 \renewcommand{\thesection}{\arabic{section}.}
\label{SecCW}

The CW-model describes a measurement of the $z$-component, $\hat{s}_z$, of a spin-$1/2$ system by a magnet. We refer the reader to \citep{OpusABN} for a detailed description of this model, here we only discuss its main points.

\begin{figure}
  \centering
  \includegraphics[scale=0.2]{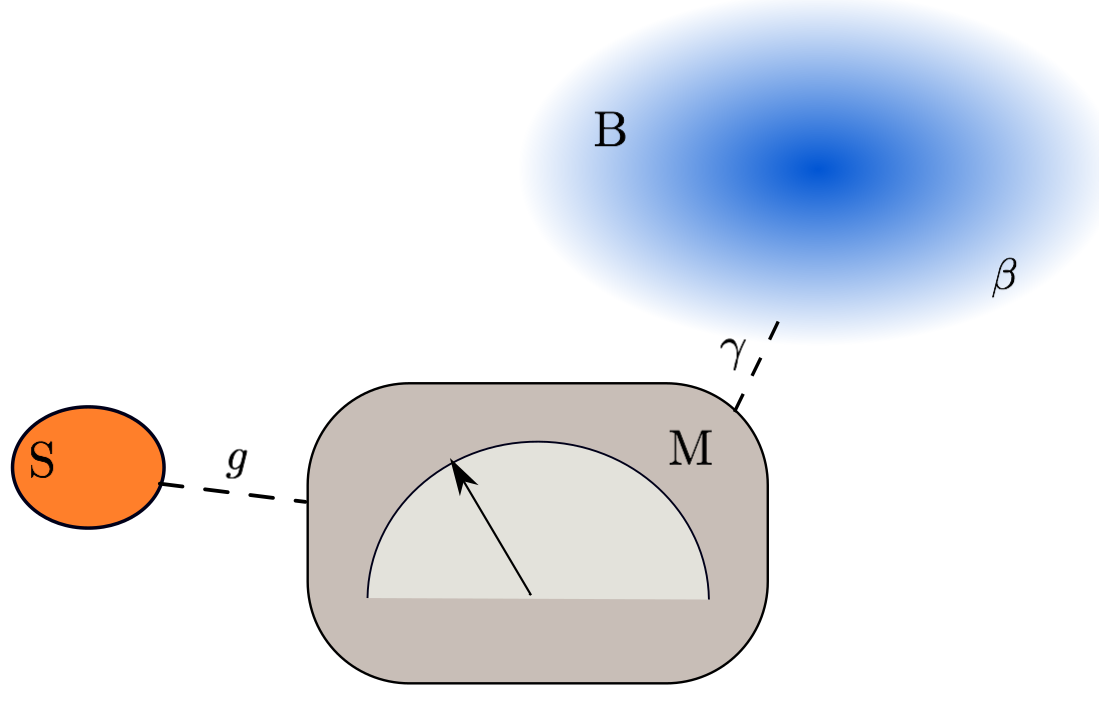}
\caption{
Schematic figure of the measurement, containing the measured spin system (S), the magnet (M), and the bath (B). The main idea of the measurement 
is that the SM interaction brakes the metastability of the initial state of M, triggering a phase transition from the initial paramagnetic state to a ferromagnetic state, 
which is driven by B. The two ferromagnetic states correspond to the pointer states of the measurement. }
\label{FigOneApp}
\end{figure}

\subsection{The Hamiltonian }
The apparatus (A) consists of a magnet (M) in contact with a thermal bath (B) -- see Fig \ref{FigOneApp}. Microscopically, M consists of a large number $N$ of spins with Pauli operators $\hat{\sigma}^{(n)}_a$ ($a=x,y,z$). The magnetization of M, 
\begin{equation}
\hat{m}=\frac{1}{N} \sum_{n=1}^N \hat{\sigma}_z^{(n)},
\end{equation}
 is the pointer variable.  For large $N$, the apparatus has macroscopic properties and the pointer turns out to take stable values that can be read off at any suitable moment.
 The magnetization is coupled to the measured variable $\hat{s}_z$ via,
\begin{equation}
\hat{H}_{\rm SA}= -N g \hat{m} \hat{s}_z.
\label{H_SA}
\end{equation} 
with $\hat{s}_z=\half \hbar \sigma_z^{(S)}$ -- throughout the article we will assume $\hbar=1$. 
This coupling selects a preferred direction $z$ for the measurement. On the other hand,  the free Hamiltonian of M is an Ising-like interaction,
\begin{equation}
\hat{H}_{\rm M}=-J_2 N \frac{\hat{m}^2}{2} - J_4 N \frac{\hat{m}^4}{4}.
\label{H_M}
\end{equation}
Let us note that this Hamiltonian plays an important role in the study of phase transitions in statistical mechanics \cite{eii,ei,eiii}. 

In turn, M is coupled to B, which is a bosonic bath made up of an infinite number of bosonic degrees of freedom with an Ohmic spectral density. Each $\hat{\sigma}^{(n)}_a$, ($a=x,y,z$), is coupled homogeneously  to the phonons of B. The full Hamiltonian, including the coupling between M and B, is presented in Appendix \ref{AppHamiltonian}.


It will be useful to decompose the magnetization as,
\begin{equation}
\hat{m}= \sum_{\{ m\}} m \hat{\Pi}_m
\end{equation}
where $\{ m \}=\{-1+2i/N \}_{i=0}^{i=N}$ are the set of eigenvalues of $\hat{m}$, and $\hat{\Pi}_m$ is a projector on the corresponding subspace. The degeneracy of each $m$ is given by,
\begin{align}
G(m) &= \Tr \left(\hat{\Pi}_m \right) = \frac{N!}{\left(\frac{N}{2}(1+m) \right)! \left(\frac{N}{2}(1-m) \right)!} 
\nonumber\\
&\approx \frac{1}{\sqrt{2\pi N}} \exp\left[ -\frac{N}{2} \left( \ln\frac{1-m^2}{4} +m \ln\frac{1+m}{1-m} \right) \right]
\label{G(m)}
\end{align}
where in the last step we used Stirling's approximation, $N! \approx \sqrt{2\pi N} (N/e)^N$, and kept only leading terms in $N$. 

\subsection{The  state}
In order to have an unbiased measurement, it is mandatory that the density matrix of A does not depend on the one of S.
The initial state of the process is taken as a product state between S, M and B,
\begin{equation}
\hat{\mathcal{D}}_0= \hat{r}_S \otimes \hat{R}_M \otimes \hat{R}_B,
\label{GlobalInitial}
\end{equation}
where the state of S is a generic spin state,
\begin{equation}
\hat{r}_S=\sum_{i,j=\{\uparrow, \downarrow\}} r_{ij} \keti \braj. 
\end{equation}
where $\ketup$, $\ketdown$ are eigenstates of $\hat{s}_z$, and  $r_{ij}= \brai \hat{r}_S \ketj$ with $i,j=\{\uparrow, \downarrow\}$. 

That the density matrix of A starts in the product state is a choice of the initial state we consider.
The state of M is a paramagnetic state, described as a maximally mixed state with zero average magnetization,
\begin{equation}
\hat{R}_M= \frac{1}{2^N} \bigotimes_{n=1}^N \mathbb{I}^{(n)}.
\label{RMinitial}
\end{equation}
This form can be achieved by putting the magnet in a strong  radiofrequency  field.
 The distribution of the magnetization in \eqref{RMinitial} is given by $P_0(m)=G(m)/2^N$, which, 
in the limit of large $N$, can be well approximated by a Gaussian distribution,
\begin{equation}
P_{0}\left(  m\right)  
\approx \sqrt{\frac{N}{2\pi}}e^{-Nm^2/2}.
\label{Pm0}
\end{equation}
 Finally, B is assumed to start out in a thermal state at temperature $\beta$,
 \begin{align}
 \hat{R}_B=\frac{e^{-\beta \hat H_B}}{\mathcal{Z}_B}.
 \end{align}
 where $H_B$ is the Hamiltonian of a bosonic bath (see Appendix \ref{AppHamiltonian} for details).
 This can be achieved by thermalizing B with a larger bath before the start of the measurement.

 When considering the evolution of \eqref{GlobalInitial} with the full Hamiltonian, the only relevant degrees of freedom are those of SM, 
 described by the reduced state $\hat{D}_{SM}(t) = \Tr_B (\hat{\mathcal{D}}(t))$. Without explicitly solving the dynamics, an important property of the evolution 
 of $\hat{D}_{SM}(t)$ is that it always admits the decomposition \cite{OpusABN} (see also Appendix \ref{AppendixForm}),
\BEQ
\label{exponeapp} \hspace{-4mm} 
\hat{D}_{SM}(t)=\sum_m \sum_{i,j=\{\uparrow, \downarrow\}} \frac{1}{G(m)}P_{ij}(m,t) \keti \braj \otimes \hat{\Pi}_m.
\EEQ
 Here, $\Puu$ ($\Pdd$) represent the conditional probability of the magnetization being equal to $m$
 given that S is pointing up (down),  whereas $\Pud$ ($\Pdu$) bring information about the coherent terms. The initial conditions are given by $P_{ij}(m,t)=r_{ij}P_{0}(m)$.  Given the decomposition \eqref{exponeapp}, the probability distribution of $m$ at time is simply given by
 \begin{equation}
 P(m,t)=\Tr(\hat{\Pi}_m \hat{D}_{SM}(t))= \Puu(m,t)+\Pdd(m,t).
 \label{P(m,t)}
 \end{equation}
 
 Note that the decomposition \eqref{exponeapp} allows us to express the state of SM, for which in principle $2^{N+1}$ degrees of freedom are required, 
 through functions that have only $\mathcal{O}(N)$ degrees of freedom - given by the possible values of $m$. 
 This property allows one to perform numerical simulations for relatively large systems, which will be particularly useful when considering two apparatuses, 
 and is also essential to analytically solve  the equations of motion.

 \subsection{Equilibrium states and free energies}
 \label{FreeEnergiesSec}

The interaction of M with the thermal bath B tends to bring M to a stable equilibrium state. Given the Hamiltonian \eqref{H_M}, there are three (locally) stable states for M: 
two stable ferromagnetic and a metastable paramagnetic state \cite{OpusABN}. In order to see that, consider the state of M  at thermal equilibrium in
 the temperature $1/\beta$ of the bath, 
\begin{align}
 \hat{R}_M^{(eq)}= \frac{e^{-\beta \hat{H}_M}}{\mathcal{Z}_M},
\end{align}
with a probability distribution given by,
\begin{align}
 P_{\rm eq}(m)= G(m)\frac{e^{-\beta H_M(m)}}{\mathcal{Z}_M}.
 \label{Pmeq}
\end{align}
One can now construct a free energy-like function $F_{\rm eq}(m)$ by inserting \eqref{G(m)} into \eqref{Pmeq}  and identifying,
\begin{align}
P_{\rm eq}(m)\equiv \frac{e^{-\beta F_{\rm eq}(m)}}{\mathcal{Z}_M},
\label{Peqm}
\end{align}
obtaining
\begin{equation}
\label{freeenergyrprm}
F_{\rm eq}(m)=H_M(m)+\frac{N}{2\beta}\left( \ln \frac{1-m^2}{4} +m \ln \frac{1+m}{1-m}\right),
\end{equation}
where we assumed $N \gg 1$ and neglected the constant term $-\ln 2\pi N /2$. 
This free energy-like function arises from the Hamiltonian $\hat{H}_M$ and the degeneracy of  $\hat{m}$ (which brings the entropic contribution). This function presents a local minimum at $m=0$ 
(paramagnetic region) and two global minima at $\pm m_{\rm F}$ with $m_{\rm F} \approx 1$ (ferromagnetic region) for low $\beta$ and $J_2<3 J_4$, 
which is  the regime where the apparatus can function for a measurement.  These (local) minima correspond to (meta)stable states of M when put in contact 
with B. As such, the probability distribution \eqref{P(m,t)} naturally evolves towards them in the course of time \cite{OpusABN}.

\begin{figure}
  \centering
  \includegraphics[scale=0.32]{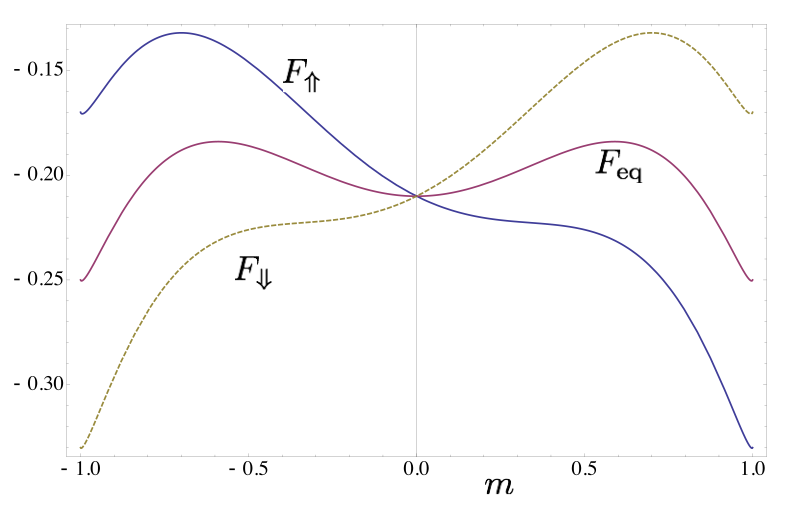}
\caption{ Free energy functions of M: (i) in absence of S ($F_{\rm eq}$), (ii) in the presence of  $\ketup$ ($F_{\Uparrow}$), and  (iii) in the presence of $\ketdown$ ($F_{\Downarrow}$). The minima of $F_{\Uparrow}$ defines the value of $m_F \approx 1$. The parameters are chosen as, $J_2=0$, $J_4=1$, $\beta=3.3$, $g=0.08$.  
}
\label{FigFreeEnergyOneApp}
\end{figure}

While \eqref{freeenergyrprm} captures the equilibrium states of M in absence of S, we are in fact interested in the joint state of SM. Let us hence consider the equilibrium state of SM, and expand it as, 
\begin{align}
\label{eqoneapp}
 \hat{D}_{SM}^{(eq)}&=\frac{e^{-\beta(\hat{H}_{SM}+\hat{H}_M)}}{\mathcal{Z}}
\nonumber\\
&=\ketup \braup \otimes \hat{R}^{(eq)}_{\Uparrow} +\ketdown \bradown \otimes \hat{R}^{(eq)}_{\Downarrow}
\end{align}
where 
\begin{align}
\hat{R}^{(eq)}_{i}=\frac{e^{-\beta(s_i N g\hat{m}+\hat{H}_M))}}{\mathcal{Z}},
\end{align}
with  $s_i=\pm 1 /2$ for $i=\{\Uparrow,\Downarrow\}$. In other words, $\hat{R}^{(eq)}_{\Uparrow}$ ($\hat{R}^{(eq)}_{\Downarrow}$) are thermal states of M with an extra positive (negative) field due to the interaction with S. In analogy with \eqref{freeenergyrprm}, we can construct free energy functions associated with (the distribution of $m$ for) those states, obtaining,
\begin{equation}
F_{\rm i}(m)=-s_i Ngm+F_{\rm eq}(m), \hspace{10mm} s_i=\pm 1/2
\label{FreeEnergyOne}
\end{equation}
with $i=\{\Uparrow,\Downarrow\}$. 
Clearly,  in absence of interaction with S, $F_{\uparrow}$ and $F_{\downarrow}$ coincide with the original $F_{\rm eq}$. Yet, the presence of $g$ breaks the symmetry $m \leftrightarrow -m$ of $F_{\rm eq}(m)$, so that the positive (negative) ferromagnetic state becomes the most stable one for $F_{\uparrow}$  ($F_{\downarrow}$).  Furthermore, if $g$ is large enough,  $F_{\rm i}$ presents no longer a local minima near $m=0$, i.e., the interaction with S breaks the metastability of the paramagnetic state and the system can be used as a measurement device that will end up in a magnetized state. These considerations are shown in Fig \ref{FigFreeEnergyOneApp}.

\subsection{Measurement process}
The joint evolution of S, M and B is captured by the following two processes: (i) A  dephasing process  due to the interaction between S and M (named truncation of the initial 
state in \cite{OpusABN}), and (ii) a decay of M from the paramagnetic state \eqref{RMinitial} towards one of the two ferromagnetic states, termed registration of the measurement.  
The former evolution takes place on a time scale $\tau_d \propto 1/g$, whereas the latter one is characterised by the time scale $\tau_{r} \propto 1/\gamma$, where 
$\sqrt{\gamma}\ll1$ is the dimensionless coupling strength of MB. 
Because of the smallness of $\gamma$ (that is, the weakness of the coupling to the bath) we have that $\tau_d \ll \tau_r$, so that the dephasing process takes place much faster. 

\subsubsection{Dephasing} 
\label{Dephasing}
Let us first focus on the interaction \eqref{H_SA}, and neglect  the presence of B, which acts on a much longer time scale.  
In this case, from \eqref{H_SA} and \eqref{exponeapp}, one obtains,
\begin{align}
\Puu(m,t)&= \ruu P_0(m) 
\nonumber\\
\Pud(m,t)&= \rud P_0(m) e^{-i2Ngmt}
\label{Pudoneappa}
\end{align}
and the other components are solved using $ \Pud(m,t)=\Pud^{*}(m,t)$ and $\ruu+\rdd=1$.
Hence, in the basis spanned by $\hat{s}_z$, the off-diagonal elements of SM gain phases whereas the diagonal elements remained unmodified. At the level of S, this leads to a decay of the off-diagonal elements, 
\begin{align}
\rud (t) &= \sum_m \Pud(m,t) 
\nonumber\\ &\approx  \rud \int P_0(m) e^{-i2Ngmt} dm =\rud e^{-t^2/\tau_d^2} 
\label{sumdephasingone}
\end{align}
 where in the second step we took the limit to the continuum (which holds strictly for $N \rightarrow \infty$) and inserted \eqref{Pm0}. 
 The decay process takes place in a time scale $\tau_{\rm d} = 1/\sqrt{2N}g$. From \eqref{Pudoneappa} and \eqref{sumdephasingone} it follows that, 
\begin{align}
\label{dephasingOne}
\langle \hat{s}_x (t)\rangle &= \langle \hat{s}_x(0)\rangle e^{-t^2/\tau_d^2} 
\nonumber\\
\langle \hat{s}_y (t)\rangle &= \langle \hat{s}_y(0)\rangle e^{-t^2/\tau_d^2} 
\nonumber\\
\langle \hat{s}_z (t)\rangle &= \langle \hat{s}_z(0)\rangle.
\end{align}
That is,  the dephasing process erases information about the non-measured observables, which is lost in the many degrees of freedom of M. 
The form (\ref{Pudoneappa}) may produce recurrencies (non-small values of $\Pud(m,t)$) at later times; these can be suppressed by a spread in the constant 
$g$ that couples to the spins of M and/or by the action of the bath. The decay (\ref{dephasingOne}) goes together with a
cascade of small correlations between the transverse components $\hat{s}_{x,y}$ of the tested spin and an arbitrary, finite number 
$\hat\sigma_z^{(i_1)}\cdots\hat\sigma_z^{(i_k)}$ of the $z$-components of the of spins of A \cite{OpusABN}:
 in this initial stage, the information coded in the transverse components is transferred to many weak multi-particle correlations in M. 
 This is still phase coherent; at a later time it may get lost by transfer to the bath (``decoherence'').

\subsubsection{Registration}

In the registration,  the information about $\hat{s}_z$ is transferred to the pointer states of M (associated with the two ferromagnetic states). 
 Without explicitly solving the dynamics, let us here give an intuition of this process using the free energies derived in \eqref{FreeEnergyOne}. Indeed,  
the free energy functions $F_{\rm i}(m)$ also bring information about the \emph{non-equilibrium} dynamics, as  $\Puu(m,t)$ and $\Pdd(m,t)$ in \eqref{exponeapp} tend to the minima of  $F_{\Uparrow}$ and $F_{\Downarrow}$, respectively \cite{OpusABN}. 

At the beginning of the measurement, we have that $\Puu(m,0)=\ruu P_M(0)$ and $\Pdd(m,0)=\rdd P_M(0)$, where $P_M(0)$ is a gaussian distribution centered in $m=0$, see 
\eqref{Pm0}. Now,  in absence of interactions, these distributions would eventually decay to an equally weighted distribution of the two ferromagnetic states (i.e. to a thermal state). 
This decay is slow because it has to overcome a free energy barrier, which demands an exponential time in $N$. However, for large enough $g$, metastability is broken, so 
that $\Puu(m,t)$ and $\Pdd(m,t)$ evolve rapidly towards the minima of $F_{\Uparrow}$ and $F_{\Downarrow}$, respectively, in a relatively short time scale of order 
$(J_2+J_4)/\gamma$. The final states of $\Puu(m,t)$  $\Pdd(m,t)$ are hence two ferromagnetic distributions peaked around $m=\pm m_F$, with $m_F\approx 1$. 
This intuitive explanation can be confirmed by explicitly solving the equations of motion of the process \cite{OpusABN}. 

 On the other hand, the off diagonal-elements $\Pdu(m,t)$
 and $\Pud(m,t)$ decay due to a decoherence effect induced by the bath, which can be anticipated by noting that the equilibrium state \eqref{eqoneapp} has no off-diagonal terms.   Putting everything together, we have, given the expansion \eqref{exponeapp}, the following form for the final state of SM,
\begin{equation}
 \hat{D}(t_F)=r_{\uparrow \uparrow} \ketup \braup \otimes \hat{R}_{\Uparrow}+r_{\downarrow \downarrow} \ketdown \bradown \otimes \hat{R}_{\Downarrow}.
 \label{finalstate}
\end{equation}
 where $\hat{R}_{\Uparrow}$, $\hat{R}_{\Downarrow}$ are the two pointer states at $\pm m_F$, i.e., $\hat{R}_{\Uparrow}\approx \hat{\Pi}_{m_F}$, $\hat{R}_{\Downarrow}\approx \hat{\Pi}_{- m_F}$ \cite{commentR}. Only in a much longer time scale, the state of SM will evolve to a thermal equilibrium state, 
 leaving ample time to read off the measurement outcome at a suitable moment. 
 
 The state \eqref{finalstate} is the expected final state of a projective measurement: With probability $r_{\uparrow \uparrow}$ ($ r_{\downarrow \downarrow}$) the state of S is projected into $\ketup$ ($\ketdown$) and the pointer state is pointing up (down). The off-diagonal elements of SM disappear, those of S due to the dephasing effect in \eqref{dephasingOne}, and those of M due to the presence of the bath (note that the equilibrium state \eqref{eqoneapp} has no off-diagonal elements).  As usual in unitary dynamics, the off-diagonal terms of the whole system SMB do not mathematically disappear but become lost at the level of SM. 
 
 Of course, as already mentioned, the justification  for \eqref{finalstate} presented here is an heuristic one, based on the free energy functions \eqref{FreeEnergyOne}. Yet, the final form \eqref{finalstate}  can be rigorously proved by solving the dynamical equations  \cite{OpusABN}.
 
 As a final remark, we note that from the expression \eqref{FreeEnergyOne}, one can find the minimum coupling $h_c$ between S and M for which the free energy barrier disappears, so that the registration process becomes possible. For $J_2=0$, one finds that,
 \begin{equation}
 h_c=\frac{T}{2} \ln\left(\frac{1+m_c}{1-m_c}\right),
 \label{h_c}
 \end{equation} 
 with $2m_c^2= 1-\sqrt{1-4T/3J_4} $ \cite{OpusABN}, where T should satisfy $T<3J_4/4$.
 Only when $g>h_c$, the apparatus will reach a ferromagnetic state, hence yield an outcome for the measurement -- in a non-exponential time in $N$.

 \renewcommand{\thesection}{\arabic{section}}
\section{Dynamics of a  joint measurement of two observables}
 \setcounter{equation}{0} \setcounter{figure}{0}
 \renewcommand{\thesection}{\arabic{section}.}
\label{Sec2app}

Let us now explore the possibility of coupling S simultaneously to two apparatuses. For that purpose we extend the previous considerations  by adding a second apparatus 
A$'$ which attempts to measure $\hat s_x$. Analogous to A, it is made up of magnet M$'$ and a bath B$'$, with parameters $ J_2', J_4', g', N'...$,  which we assume to 
have the same order than those of A, and an internal Hamiltonian $H_{\rm M'}$ analoguous to (\ref{H_M}). The initial state of M'  is also a paramagnetic state and it has two 
pointer states corresponding to the two ferromagnetic states. Therefore,  there are four possible pointer states, and hence four outcomes of the measurement $\{--,-+,+-,++ \}$. 
We aim to extract information about the expectation value of the two measured observables, $ \hat{s}_z$ and $ \hat{s}_x$, from such outcomes.

While, for convenience, we assume that each magnet interacts with its own bath,   both baths must have the same temperature $1/\beta$, so that no heat currents are present. The magnets are expected to eventually equilibrate to the thermal state \eqref{eqtwoapp}. In other words, the two magnets share a common thermal environment, which interacts locally and independently with each magnet.    

In the following we describe the main characteristics of the dynamics of this process. We always try to keep  the analogy with the considerations for the single-apparatus measurement as close as possible.  
%

\begin{figure*}
  \includegraphics[width=\textwidth]{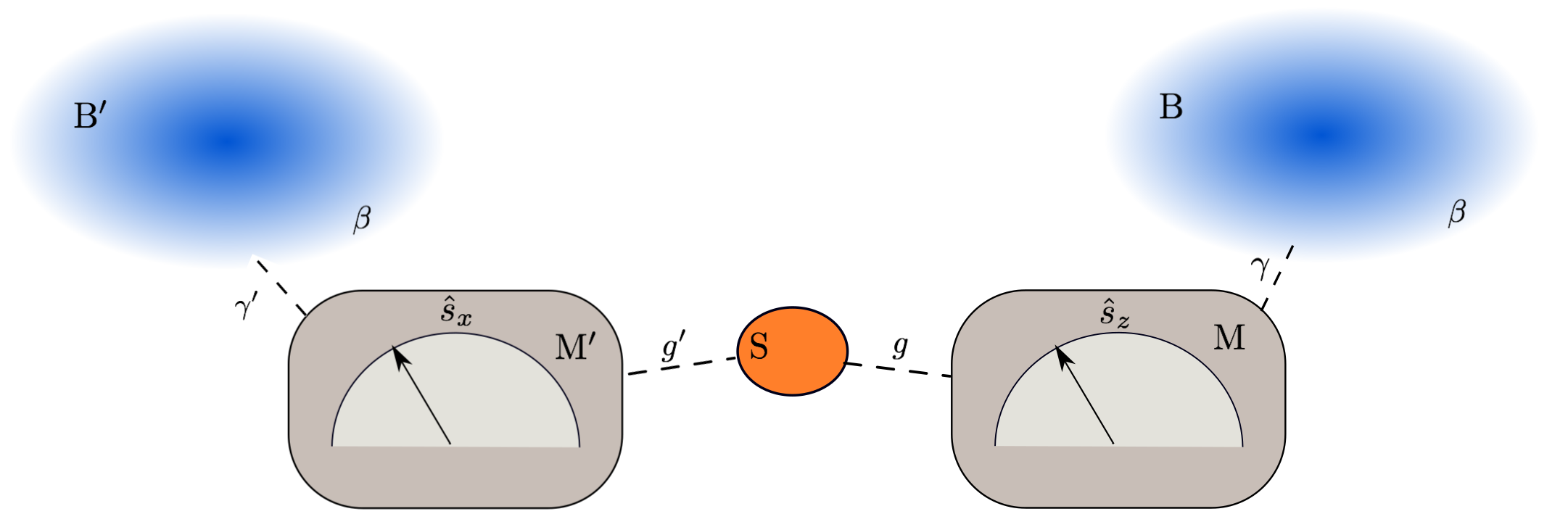}
  \caption{Schematic figure of the simultaneous measurement. The system S simultaneously interacts with M and M' through the $z$ and $x$ spin component, respectively. }
\end{figure*}

\subsection{The Hamiltonian}

Let S interact simultaneously with both apparatuses through,
\BEQ \hspace{-7mm}
\hat{H}_{\rm SMM'}=-Ng \hat{s}_z \otimes \hat{m} \otimes \mathbb{I}_{M'} -N'g' \hat{s}_x \otimes \mathbb{I}_M \otimes \hat{m}',
\label{inttwo}
\EEQ
where $\mathbb{I}_M$ and $\mathbb{I}_{M'}$ are the identity in the Hilbert space of M and M$'$, respectively. 
In analogy with Eq. \eqref{H_SA}, note that A is attempting to measure $ \hat{s}_z$ and A' $ \hat{s}_x$. By expanding the magnetizations as
\begin{align}
\hat{m}&= \sum_{\{ m\}} m \hat{\Pi}_m \nonumber\\
\hat{m}'&= \sum_{\{ m'\}} m' \hat{\Pi}_{m'},
\label{expansionm}
\end{align}  
where $\hat{\Pi}_i$  ($\hat{\Pi}_j'$) is a projector in the subspace spanned by the eigenvectors with eigenvalue $m$ ($m'$), we can write $\hat{H}_{\rm SMM'}$ as,
\begin{align}
\hat{H}_{\rm SMM'}&= -\sum_{m,m'} (Ng m \hat{s}_z+N'g' m' \hat{s}_x)\otimes  \hat{\Pi}_m \otimes  \hat{\Pi}_{m'} 
\nonumber\\
&=-\sum_{m,m'} w(m,m') \hat{s}_{\bf u} (m,m') \otimes  \hat{\Pi}_{m} \otimes \hat{\Pi}_{m'} 
\end{align}
where we have defined a modulus $w$ and unit vector ${\bf u}$,
\begin{align}
&w(m,m')=\sqrt{(Ngm)^2+(N'g'm')^2}, 
\label{definitionww}
\\
&{\bf u}(m,m')= \frac{Ngm {\bf z}+ N'g'm' {\bf x}}{w(m,m')},
\label{definitionu}
\end{align}
and the spin projection 
\BEQ
 &\hat{s}_{\bf u}(m,m')={\bf u}(m,m')\cdot {\bf \hat{s}}. 
 \label{defsu}
\EEQ
Hence, we see that S effectively acts on both apparatuses as a global field $w(m,m')$ in the direction $\bf{u}$$(m,m')$. Note that, for every value of the magnetization of the apparatuses, the field acts with a different strength and direction. 

In what follows, to avoid cumbersome expressions, we will sometimes not write explicitly the dependence on $(m,m')$ of ${\bf u}(m,m')$ and $w(m,m')$. 
We will also denote ${\bf u}=(u_x,0,u_z)$,  and define a direction ${\bf v}$ in the $x-z$ plane perpendicular to ${\bf u}$, that is,
${\bf v}(m,m')=(u_z ,0, - u_x)$, so that  the spin projected on it reads $\hat{s}_{\bf v}=u_z \hat{s}_x - u_x \hat{s}_z$. 
It is useful also to introduce the states $\ketupu$, $\ketdownu$, which are eigenvectors of $s_{\bf u}$,
\begin{equation}
\hat{s}_{\bf u} \ketupu = \frac{1}{2}\ketupu, \quad \quad \hat{s}_{\bf u} \ketdownu = -\frac{1}{2}\ketdownu. 
\end{equation}

Finally, recall that every magnet has an internal Hamiltonian given by (\ref{H_M}) -- where in order to obtain the internal Hamiltonian of M$'$ one should 
replace $J_2 \mapsto J_2'$, $J_4 \mapsto J_4'$,  etc. Furthermore, every magnet interacts with its own bosonic bath at temperature $1/\beta$ and $1/\beta'$,
respectively, see the Appendix for the explicit form of the interaction and the internal Hamiltonian of the baths. The strength of the interaction MB, MB$'$ is 
given by $\gamma, \gamma'$, respectively. It is satisfied that $g,g' \gg \gamma, \gamma'$

\subsection{The state}
The initial state is now taken as a product state between all different elements of the measurement,
\begin{equation}
\mathcal{D}= \hat{r}_S \otimes \hat{R}_M \otimes \hat{R}_B  \otimes \hat{R}_M' \otimes \hat{R}_B',
\end{equation}
The initial state of both $\hat{R}_M$ and $\hat{R}_M'$ is the paramagnetic state \eqref{RMinitial}. 
The state of SMM'  in the course of time can always be decomposed as (see Appendix \ref{AppendixForm}),
\begin{align}
&\hat{D}_{ SMM'}(t)= \Tr_{B,B'} \mathcal{D}(t)
\nonumber\\
&\hspace{10mm}=\hspace{-3mm}\sum_{i,j=\{\uparrow, \downarrow\}} \sum_{m,m'} \frac{P_{ij}^{(\bf{u})}(m',m',t)}{G(m)G(m')} \ketiu \braju  \otimes \hat{\Pi}_{m}  \otimes \hat{\Pi}_{m'}  . 
\label{DMM}
\end{align}
In this decomposition, $\Puu^{(\bf u)}$ ($\Pdd^{(\bf u)}$) represent the conditional probability of the magnetizations given that S is in state $\ketupu$, $\ketdownu$, respectively. 
At any moment in time, the probability distribution of the magnetizations can be then expressed as,
\begin{align}
P(m,m',t)&= \Tr \left(\Pi_m \Pi_{m'} \hat{D}_{ SMM'}(t) \right) 
\nonumber\\&= \Puu^{(\bf u)}(m,m',t) +\Pdd^{(\bf u)}(m,m',t).
\end{align}
We recall that ${\bf u}={\bf u}(m,m')$ is defined at given values of $m$  and $m'$ and  that $\uparrow$ and $\downarrow$ are defined with respect to this direction.

The initial conditions for \eqref{DMM} read, 
\begin{equation}
P_{ij}^{(\bf{u})}(m,m',0 )=P_0(m) P_0'(m')r_{ij}^{(\bf{u})} , 
\end{equation}
where $r_{ij}^{(\bf{u})}=\braiu \hat{r}_S\ketju$  with $i,j=\{\uparrow, \downarrow\}$, $\hat{r}_S$ is the initial state of S,
and $P_0'(m')$ is like $P_0(m)$ but with modified parameters. 


\subsection{Dephasing} 

Let us now neglect the interaction between M, M' and B,B', as $g,g' \gg \gamma, \gamma'$; and focus on the evolution of SMM' under \eqref{inttwo}. Using the decomposition \eqref{DMM}, we find, 
\begin{align}
 P_{\uparrow\downarrow}^{(\bf{u})}(m,m',t)&=e^{i\omega(m,m')t}P(m) P(m') r_{\uparrow \downarrow}^{(\bf{u})} 
 \nonumber\\
  P_{\uparrow\uparrow}^{(\bf{u})}(m,m',t)&=P(m) P(m') r_{\uparrow \uparrow}^{(\bf{u})} ,
 \label{Puu}
\end{align}
while the other components can be determined using $ P_{\uparrow,\downarrow}^{(\bf{u})}(m,m',t)=P_{\uparrow,\downarrow}^{(\bf{u})}(m,m',t)^{*}$ and $r_{\uparrow \uparrow}^{(\bf{u})} +r_{\downarrow \downarrow}^{(\bf{u})} =1$. From this solution we can work out the evolution of S by summing over $\{ m,m' \}$, which is done in detail in Appendix \ref{DynamicsTwoApp}. Assuming $N,N' \gg 1$, we find that,
\begin{align}
\langle &\hat{s}_y (t) \rangle \xrightarrow{t \gg \tau_d} 0
\nonumber\\
\langle &\hat{s}_x (t) \rangle \xrightarrow{t \gg \tau_d}  \langle \hat{s}_x (0)\rangle  \frac{\sqrt{N'}g'}{\sqrt{N}g+\sqrt{N'}g'}
\nonumber\\
\langle &\hat{s}_z (t) \rangle \xrightarrow{t \gg \tau_d}  \langle \hat{s}_z (0)\rangle  \frac{\sqrt{N}g}{\sqrt{N'}g'+\sqrt{N}g}
 \label{sxszdecay}
\end{align}
If the two apparatuses are identical, the decay takes place in a time scale $\tau_{\rm d} = 1/\sqrt{2N}g$ --otherwise, the stronger coupling fixes the time scale.   

This partial dephasing can be intuitively understood from  \eqref{Puu}. For every value of $( m,m' )$, the diagonal elements, in the basis spanned by 
$\ketup_{\bf u}, \ketdown_{\bf u}$, remain preserved in time, whereas the off diagonal elements gain phases. When averaged over all values of $( m,m' )$, 
those phases lead to dephasing, i.e., disappearance of the off-diagonal elements of S (see the Appendix \ref{DynamicsTwoApp} for detailed calculations). Since the preferred direction ${\bf u}$ 
is always a combination of $x$ and $z$,   $ \hat{s}_{\bf u} = u_x \hat{s}_x +u_z \hat{s}_z$, with $u_x^2+u_z^2=1$, we finally obtain the partial dephasing in \eqref{sxszdecay}. 
 
 We also note from expressions \eqref{sxszdecay} that  information about the initial state of the measured components, $x$ and $z$, is partly lost. The exact tradeoff is determined by the coupling strengths of S with each apparatus. The stronger the interaction to one apparatus, the more information is kept about the corresponding observable. In particular, if we take $g' \to 0$, we obtain that  $\langle \hat{s}_z (t) \rangle \xrightarrow{t\rightarrow \infty} \langle \hat{s}_z (0)\rangle$ and $\langle \hat{s}_x (t) \rangle \xrightarrow{t\rightarrow \infty}0$, hence recovering  the results known for the single-apparatus case.  
 When the apparatuses are identical, $\langle \hat{s}_x  \rangle$ and $\langle \hat{s}_z  \rangle $ will both lose a factor $2$, while  $\langle \hat{s}_y  \rangle $
 is completely lost.
In expressions \eqref{sxszdecay} we thus observe the first signature of a competition between the two apparatuses, as well as non-ideality of this process.

\begin{figure*}
  \includegraphics[width=\textwidth]{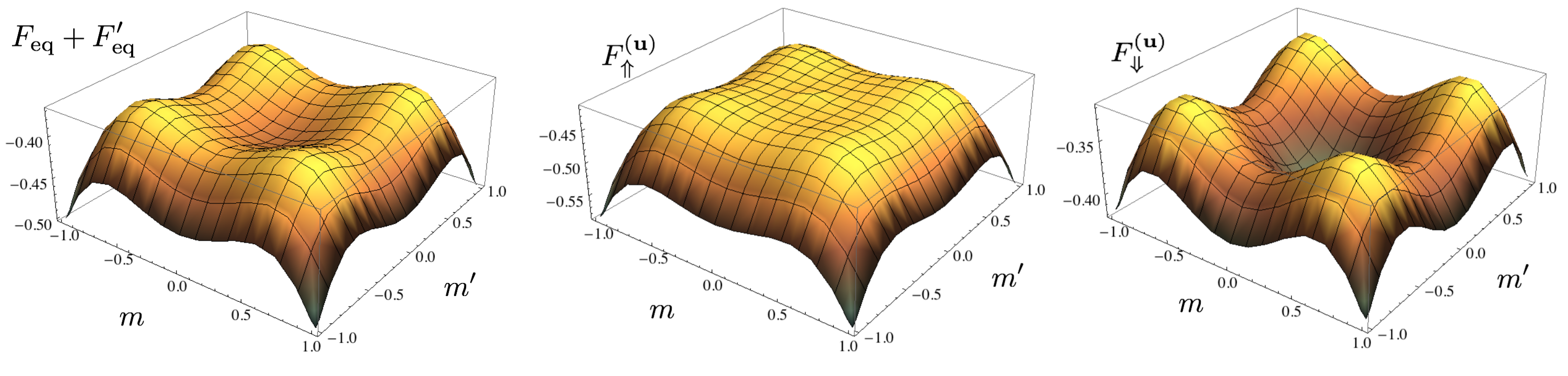}
\caption{Free energy functions of MM': (i) in the absence of S ,$F_{\rm eq}+F_{\rm eq}'$, (ii) in the presence of  $\ketupu$, $F_{\Uparrow}^{(\bf u)}$, and  (iii) in the presence of $\ketdownu$, $F_{\Downarrow}^{(\bf u)}$.}
\label{FreeEnergyTwo}
\end{figure*}


%

\subsection{Registration}

Consider now the registration process, which involves the combined effect of S with M, M' and B, B'. In Appendix \ref{DynamicsTwoApp}, we work out the corresponding equations of motion. By tracing out B and B', and taking standard approximations in open quantum systems owing to the weak coupling between MB and M$'$B$'$ \cite{breuer}, we obtain a set of equations for the evolution of $ P_{ij}^{(\bf{u})}(m,m',t)$ in \eqref{DMM}.  
The resulting equations of motion become notably complex and are given in Appendix \ref{DynamicsTwoApp}. 
Here, instead, we  describe the main features of the dynamics and the form of the final state. For that, in analogy with our considerations for one apparatus, we construct free energy functions from which the final equilibrium states and important properties of the dynamics can be inferred. The discussion is complemented with numerical simulations of the dynamics, obtained through the equations of motion derived in Appendix \ref{DynamicsTwoApp}. 



\subsubsection{Free energy function}
\label{secFreeEnergyTwoApp}

In analogy with \eqref{eqoneapp}, let us expand the thermal equilibrium state of SMM' as,
\begin{widetext}
\begin{equation}
\begin{aligned}
\label{eqtwoapp}
 \hat{D}_{ SMM'}^{(eq)}=&\frac{e^{-\beta(\hat{H}_{SMM'}+\hat{H}_M+H_{M'})}}{\mathcal{Z}}
= \frac{1}{\mathcal{Z}}\sum_{m,m'} e^{-\beta(- w \hat{s}_{\bf u}+H_M(m)+H_{M'}(m'))  }\otimes   \hat{\Pi}_{m} \otimes \hat{\Pi}_{m'}   
\\
= &\sum_{m,m'} \left(e^{-\beta H_{\Uparrow}^{(\bf u)}(m,m')  }  \ketupu \braupu+ e^{-\beta H_{\Downarrow}^{(\bf u)}(m,m')  }  \ketdownu \bradownu \right) \otimes  \hat{\Pi}_{m} \otimes \hat{\Pi}_{m'}  
\end{aligned}
\end{equation}
\end{widetext}
where we have introduced, $H_{i}^{(\bf u)}(m,m')=s_i  w(m,m')+ H_M(m)+H_{M'}(m')$ with  $i=\{ \! \Uparrow, \Downarrow\}$ and $s_i=\mp\, 1/2$. Now, proceeding in close analogy with the derivation for \eqref{freeenergyrprm}, we also construct the free energies,
 \begin{equation}
 F_{i}^{(\bf u)}(m,m')=-s_i w(m,m')  +F_{\rm eq}(m) + F_{\rm eq}'(m'),
 \label{FreeEnergyTwoApp}
 \end{equation}
 where $i=\{\Uparrow, \Downarrow \! \}$, $s_i=\pm 1/2$, and $F_{\rm eq}(m) $,  $F_{\rm eq}'(m')$ can be obtained from \eqref{freeenergyrprm}.
These free energy functions are associated with the states of MM$'$ in thermal equilibrium with the baths under the effect of S when pointing either at $+{\bf u}$ or at $-{\bf u}$ direction.  In absence of interaction with S, the $F_{i}^{(\bf u)}$ coincide and present nine (local) minima corresponding to $(0,0)$, $(0,\pm m_{\rm F})$, $(\pm m_{\rm F},0)$ and $(\pm m_{\rm F},\pm m_{\rm F} )$ in the space of $(m,m')$. As we increase $w$, the local minima in $(0,0)$ of $F_{\Uparrow}^{(\bf u)}$ loses stability whereas all ferromagnetic states become more stable. The opposite effect occurs for $F_{\Downarrow}^{(\bf u)}$: its paramagnetic point becomes more stable. The different free energies are plotted in Fig. \ref{FreeEnergyTwo}.

\begin{figure}
  \centering
  \includegraphics[scale=0.17]{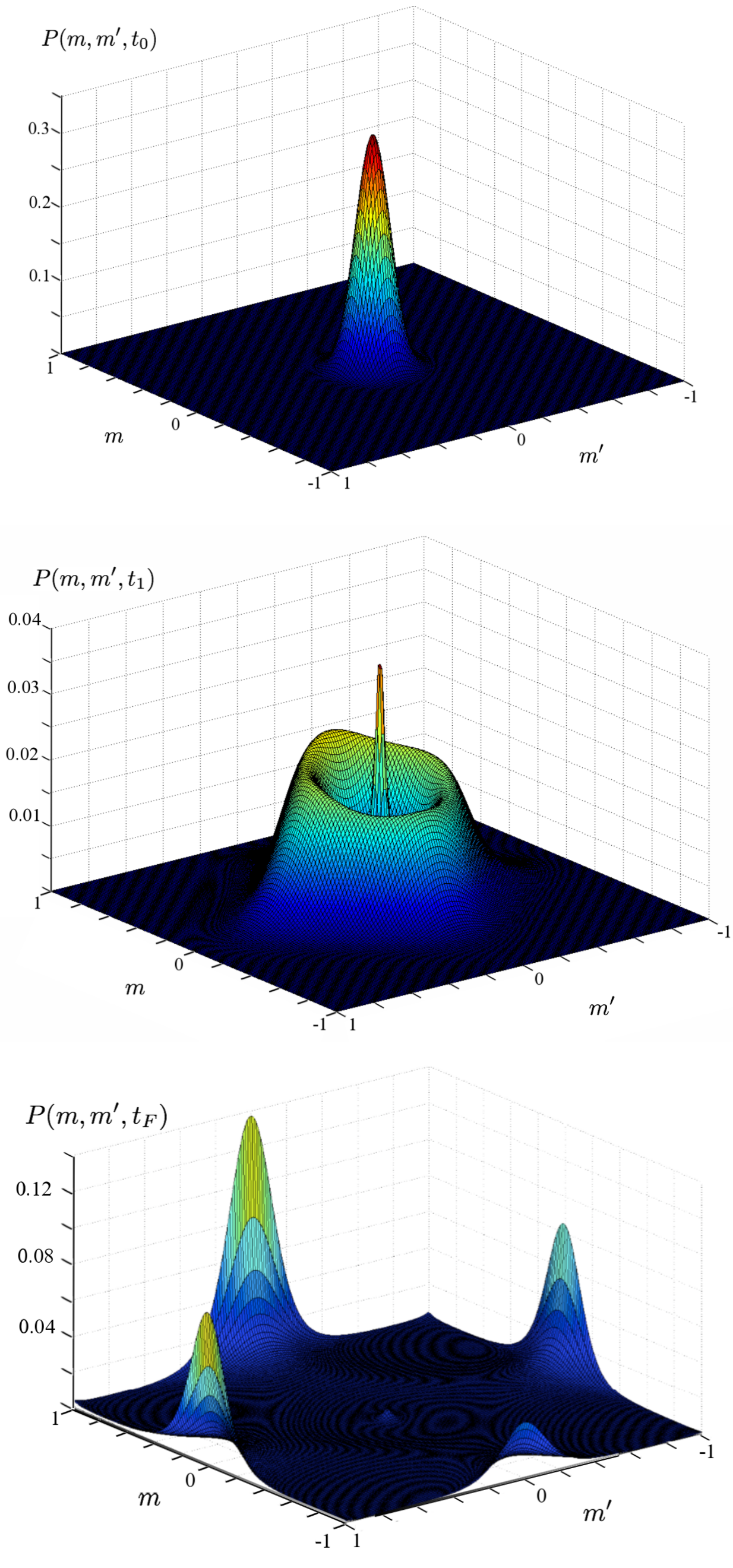}
\caption{Plot of $P(m,m',t)$ for three different times: $t=0, 6\tau, 12\tau$ (from top to bottom) with $\tau =  1/\gamma J$. 
We take: $N = 161, J_4=J, g = 0.1J, 1/\beta = 0.2 , J_2=0$, and the initial condition $\langle s^{z}(0) \rangle = 1$. 
The results are obtained by solving numerically the equations of motion \eqref{secondeqmotiontwoobservables} derived in Appendix \ref{DynamicsTwoApp}.   
Notice that at the end of the measurement only one of the two apparatuses registers a result, to be expected as the interaction satisfies $h_c<g<h_d$.
 }
\label{Fi:plotPIIdfas}
\end{figure}

\subsubsection{Dynamics of the process}
The free energy functions $F_{i}^{(\bf u)}$  are also rather useful to qualitatively describe the evolution of $D_{SMM'}$ in \eqref{DMM}, as we can associate 
$F_{\Uparrow}^{(\bf u)}$ to $P_{\uparrow\uparrow}^{(\bf{u})}(m,m',t)$, and similarly $F_{\Downarrow}^{(\bf u)}$  to $P_{\downarrow\downarrow}^{(\bf{u})}(m,m',t)$. 
Each distribution evolves to the minima of each associated free energy. This is well illustrated in Fig. \ref{Fi:plotPIIdfas}, where we numerically solve the equations 
of motion derived in Appendix \ref{DynamicsTwoApp}, obtaining $P(m,m',t)$. Initially both magnets are set in paramagnetic states, so that $P(m,m',t)$  is a two dimensional gaussian distribution. 
Notice then how $P(m,m',t)$ splits into two distributions: one, $ P_{\uparrow\uparrow}^{(\bf{u})}$, is moving towards ferromagnetic states and the other, 
$P_{\downarrow\downarrow}^{(\bf{u})}$, moves even further  towards the center. This nicely agrees with our considerations, as $F_{\Downarrow}^{(\bf u)}$ has a global 
minimum in the centre since $w(m,m')\sim\textrm{max}(m,m')>0$ there, see (\ref{definitionww}); whereas  the minima of $F_{\Uparrow}^{(\bf u)}$ are the four ferromagnetic states 
(see Fig.  \ref{FreeEnergyTwo}). On the other hand, we also  observe how $P_{\downarrow\downarrow}^{(\bf{u})}$ loses its weight until its complete disappearance, 
it being transferred to $ P_{\uparrow\uparrow}^{(\bf{u})}$. This cannot be explained from the free energy functions, and is a consequence of the fact that the 
equations of motion for $ P_{\uparrow\uparrow}^{(\bf{u})}$ and $P_{\downarrow\downarrow}^{(\bf{u})}$ are coupled. The dynamical transfer from 
$P_{\downarrow\downarrow}^{(\bf{u})}$  to $ P_{\uparrow\uparrow}^{(\bf u)}$ is discussed analytically from a simplified version of the equations of motion in Appendix \ref{DynamicsTwoApp}.

\begin{figure}
  \centering
  \includegraphics[scale=0.17]{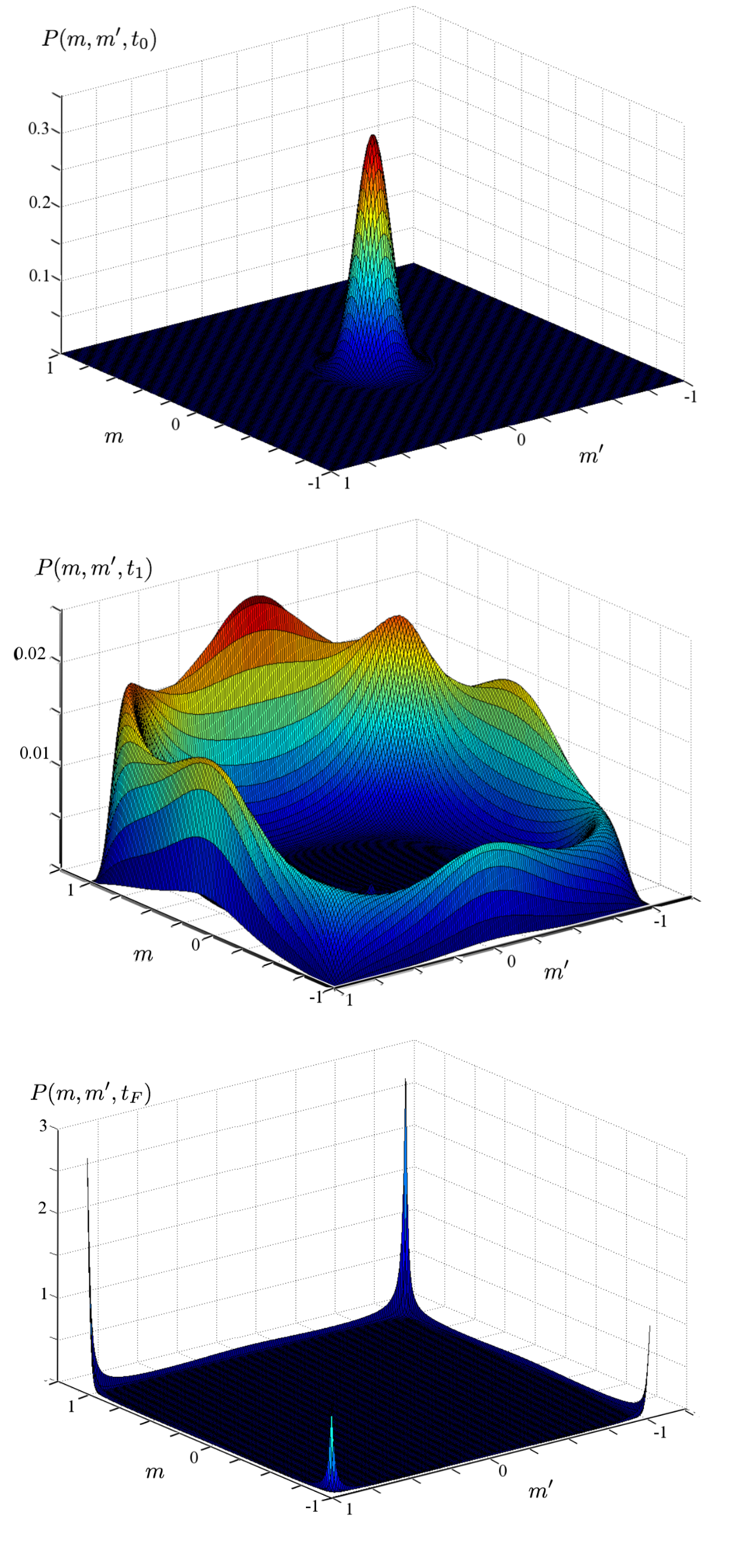}
\caption{Plot of $P(m,m',t)$ for three different times: $t=0, 6\tau, 8\tau$ (from top to bottom) with $\tau =  1/\gamma J$. We take exactly the same conditions as in Fig. \ref{Fi:plotPIIdfasII}, except for the interaction, which is increased to $g=0.4J$. In this case $g\geq h_d$, and hence both magnets can register results from the measurement.}
\label{Fi:plotPIIdfasII}
\end{figure}

Finally, we note that the off-diagonal terms in  \eqref{DMM}, given by  $ P_{\uparrow\downarrow}^{(\bf{u})}$ and $P_{\downarrow\uparrow}^{(\bf{u})}$, disappear due to a decoherence effect induced by the bath. This can be anticipated on the bases of the equilibrium form \eqref{eqtwoapp}, which indeed contains no such off-diagonal terms. Again, these considerations are corroborated by the solution of the equations of motion derived in Appendix \ref{DynamicsTwoApp}.   Finally, note that this decoherence process enhances the decay induced by the degrees of freedom of the magnet in \eqref{sxszdecay}.

\subsubsection{When do both apparatuses register a result?}
The free energy \eqref{FreeEnergyTwoApp} also allows us to find the minimal coupling $h_d$ necessary for the joint measurement to yield registration by both apparatuses. 
That is to say, the minimum value of $g$, $g'$ such that $F_{\Uparrow}^{(\bf u)}$ presents no free energy barriers ($F_{\Downarrow}^{(\bf u)}$ always presents barriers), 
such that the initial paramagnetic state centered at $(m,m')=(0,0)$ can reach one of  the four pointer states at $(\pm m_{\rm F},\pm m_{\rm F} )$ in a time 
non-exponential in $N,N'$. In Appendix \ref{AppMinimalCoupling} we derive the corresponding conditions, which allows us to find $h_d$. For $g=g'$, the result simplifies to,
\begin{equation}
h_d=2  \max (T-J_2,T-J_2' ). 
\label{h_d}
\end{equation}

\begin{figure}
  \centering
  \includegraphics[scale=0.295]{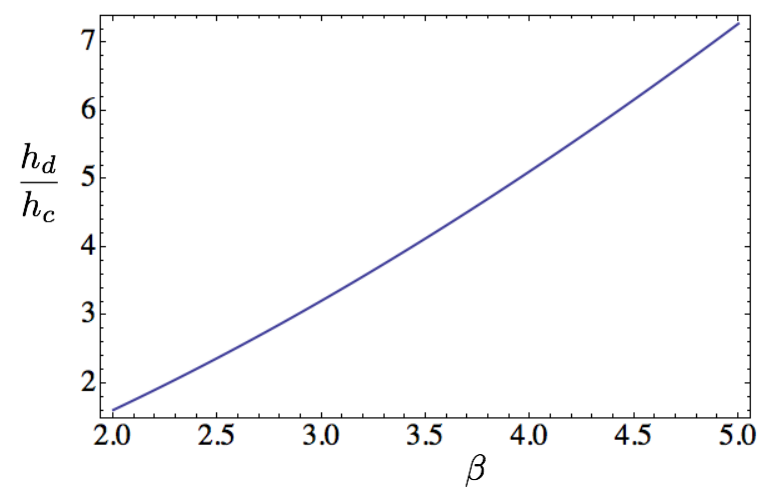}
\caption{Ratio $h_d/h_c$ as a function of $\beta$. In the Figure we take $J_2=0$ and $J_4=1$.}
\label{Fi:hc/hd}
\end{figure}

In Fig. \ref{Fi:hc/hd} we compare $h_d$ with $h_c$, the minimal coupling required for one apparatus to yield a result given in \eqref{h_c}. The Figure clearly shows how $h_d$ is considerable larger than $h_c$, the difference becoming especially noticeable for small temperatures.    Hence, we can identify different regimes depending on the value of the couplings \cite{footnoteg}, assuming $g=g'$,
\begin{itemize}
 \item for $g<h_c$ neither of the apparatuses registers a result 
 \item for $h_c<g<h_d$, only one apparatus registers a result  
\item for $g>h_d$, both apparatuses register a result  
\end{itemize}
in a time non-exponential in $N,N'$.
The different regimes are illustrated in Figs. \ref{Fi:plotPIIdfas} and \ref{Fi:plotPIIdfasII}, where we plot $P(m,m',t)$ for two different couplings to the apparatuses. 
In Fig. \ref{Fi:plotPIIdfas}, the interaction is chosen to satisfy $h_c<g<h_d$, together with $g'=g$, and indeed only one apparatus reaches a ferromagnetic state 
while the other stays practically at its initial state; after decoupling the system S from the apparatuses A and A', it will return to its parametric state.
In Fig. \ref{Fi:plotPIIdfasII}, the coupling is larger than $h_d$, and then both magnets reach ferromagnetic states, so that at the end of the measurement 
$P(m,m',t)$ is peaked at the four possible ferromagnetic states, associated with the four outcomes of the measurement.

Also during the registration processes a competition between the two apparatuses takes place. Indeed, notice that the action of S on the apparatuses is captured by, 
\begin{equation}
H_{SMM'}(m,m')=w (u_x \hat{s}_x + u_z \hat{s}_z),
\label{HSMM'}
\end{equation}
 where the first term couples S to A' and the second one S to A. Assume that $N=N'$ and $g>g'$, so that $u_z>u_x$ and hence initially S couples more strongly to
  M than to M'.  Therefore, we expect that the magnetization of M, $m$, will increase faster than $m'$. In this case, $u_z$ becomes even more dominant with respect 
  to $u_x$, thus penalizing the interaction of S with A'. Hence, as we have anticipated already, one apparatus can prevent the other one from registering the result. 
  Only when both apparatuses have an enough comparable interaction strength
 and can effectively influence each other through the measured spin S,
 each of them can register outcomes for the measurement, as in Fig \ref{Fi:plotPIIdfasII}.  


\subsubsection{The final state}

If the interaction with both apparatuses is strong enough, i.e. $g,g'>h_d$,  both apparatuses evolve to ferromagnetic states $\hat{R}_{\Uparrow}$,  $\hat{R}_{\Downarrow}$ 
with a magnetization peaked at $\pm m_F$, with $m_F\approx 1$ --see e.g. Fig. \eqref{Fi:plotPIIdfasII}. 
After this has been achieved, the couplings $g$  and $g'$ between S and the apparatuses are cut, after which the states of A and A' relax to their $g=0$ and $g'=0$ states,
respectively. 
Let us assume, for simplicity, that $\hat{R}_{\Uparrow} \approx  \hat{\Pi}_{ m_F}/G(m_F)$ and $\hat{R}_{\Downarrow} \approx  \hat{\Pi}_{-m_F}/G(m_F)$, which is only strictly true in the limit $N \rightarrow \infty$.
Since $m,m'$ are peaked at $\pm m_F$, we have that the direction ${\bf u}$ can only take four possible values at the end of the measurement, given by,
\begin{align}
{\bf u}^{(\epsilon \epsilon')} &\equiv  \frac{ \left(\epsilon' N'g' ,0,\epsilon Ng  \right)}{\sqrt{(Ng)^2+(N'g')^2}}, \hspace{10mm}\epsilon, \epsilon' = \pm 
\label{defuf}
\end{align}
which is found by inserting $\pm m_F$ into \eqref{definitionu}.  Let us also define  the states  $\ketupee$, with $\epsilon, \epsilon' = \pm $,
which are states pointing at the  ${\bf u}^{(\epsilon \epsilon')}$ direction, i.e., 
\begin{align}
\left(u_z^{(\epsilon \epsilon')} \hat{s}_z+ u_x^{(\epsilon \epsilon')} \hat{s}_x \right) \ketupee =\frac{1}{2} \ketupee.
\label{defketupee}
\end{align}
From   \eqref{DMM}, we notice that these are the possible states of S at the end of the measurement.  

Let us look in detail at \eqref{DMM} for the final state. We have already argued that the diagonal terms $P_{\uparrow \uparrow}^{({\bf u})}$ 
tend to four peaked distributions corresponding to the four ferromagnetic states.  
On the other hand,  the off-diagonal terms $ P_{\uparrow\downarrow}^{(\bf{u})}$ 
and $P_{\downarrow\uparrow}^{(\bf{u})}$ disappear due to the rapid oscillations and the interaction of the bath; and so does  $P_{\downarrow\downarrow}^{(\bf{u})}$ 
through a mechanism  discussed in Appendix \ref{DynamicsTwoApp}. Putting everything together, and using the expansion \eqref{DMM},
we can write the final state after the registration as,
\begin{widetext}
\begin{align}
\hat{D}_{\mbox{\tiny SMM'}}(t_F) \hspace{-1mm}
&= \hspace{-1mm}  \Ppp \ketupepp \braupepp \otimes    \hat{R}_{\Uparrow} \otimes \hat{R}_{\Uparrow}' 
+ \Ppm\otimes \ketupepm \braupepm    \otimes \hat{R}_{\Uparrow} \otimes \hat{R}_{\Downarrow}' 
+ \Pmp \ketupemp \braupemp     \otimes \hat{R}_{\Downarrow} \otimes \hat{R}_{\Uparrow}' \nn\\
&+ \Pmm \ketupemm \braupemm   \otimes  \hat{R}_{\Downarrow} \otimes \hat{R}_{\Downarrow}' .
\label{form2measurements}
\end{align}
\end{widetext}
where $p_{\epsilon \epsilon'}$ are the weights of each peak, $p_{\epsilon,\epsilon'}= P(\epsilon m_F, \epsilon' m'_F, \tau_f)$, where $\tau_f$ is a time where the measurement has been registered ($\tau_f \propto 1/\gamma J $).

\begin{figure}
\vspace{5mm}
  \centering
  \includegraphics[scale=0.22]{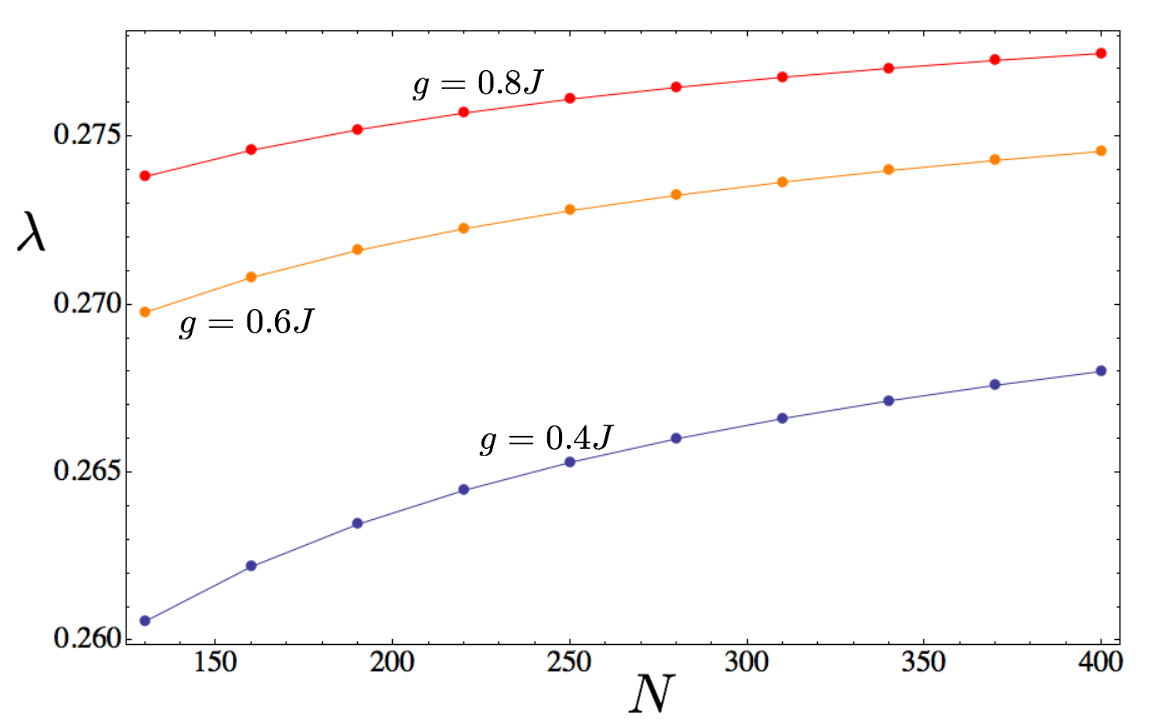}
\caption{Numerical estimates of $\lambda$ using the equations of motion \eqref{secondeqmotiontwoobservables} ,  in the case of two identical apparatuses. The parameters are $J_4=J, J_2=0, 1/\beta = 0.2J$ and $\lambda$ is evaluated at $t_{f} = 10\tau$ with $\tau=1/\lambda$, a time for which the measurement is finalised.}
\label{Lambda}
\end{figure}

Expression \eqref{form2measurements} involves  a convex sum of four independent terms, each of them corresponding to a different outcome of the experiment $\{++,+-,-+,-- \}$. 
The probability of each outcome is given by $\Pee$, with $\epsilon, \epsilon' = \pm $. In general, those $\Pee$ depend on the initial state of S and 
{\it also on the parameters of the apparatuses}, clearly expressing the non-ideality of the measurement. 
Let us discuss the dependence of such weights on the initial conditions of S following \cite{OpusABN}. The equations of motion for $P(m,m',t)$ derived in appendix D involve distributions whose initial conditions depend on S through $\langle \hat{s}_u (0) \rangle $, which is a linear combination of $\langle \hat{s}_x(0) \rangle$ and $\langle \hat{s}_z(0) \rangle$. Linearity of the equations of motion then implies that the final state should also be a linear combination of them. On the other hand, if $\langle \hat{s}_x(0) \rangle=\langle \hat{s}_z(0) \rangle=0$, then we have $\Pee=1/4$ due to the symmetries $m \leftrightarrow-m$ and $m' \leftrightarrow -m'$. These symmetries also imply that, for $\langle s_x \rangle=0$, then $p_{\epsilon +}= p_{\epsilon -}$; and similarly for  for $\langle s_z \rangle=0$, then $p_{+ \epsilon}= p_{- \epsilon}$. Putting everything together, we can write
\begin{equation}
\Pee =\frac{1}{4}(1+ \epsilon \lambda \langle \hat{s}_z (0) \rangle+  \epsilon' \lambda' \langle \hat{s}_x (0) \rangle)
\label{Peedeff}
\end{equation}
where, due to positivity, $\lambda$ and $\lambda'$ satisfy  $\{\lambda,\lambda'\}\in [0,1]$. 
Notice that the linearity imposes absence of $\epsilon\times\epsilon' $ terms in (\ref{Peedeff}).
Determining the specific form of   $\lambda$  and $\lambda'$  requires in general solving  the dynamics. In Fig. \ref{Lambda}, 
we determine them numerically for a joint measurement with two identical apparatuses. Relatively large values of $\lambda=\lambda'$ are 
experimentally preferable because they expose less noise.


Finally, let us write the states \eqref{defketupee} explicitly, finding,
\begin{align}
\ketupepp & \propto u_x^{(++)} \ketup + \left(1-u_z^{(++)} \right) \ketdown
\nonumber\\
\ketupepm& \propto u_x^{(++)} \ketup - \left(1-u_z^{(++)} \right) \ketdown 
\nonumber\\ 
\ketupemp& \propto u_x^{(++)} \ketup + \left(1+u_z^{(++)} \right) \ketdown 
\nonumber\\
\ketupemm& \propto u_x^{(++)} \ketup - \left(1+u_z^{(++)} \right) \ketdown
\end{align}
 Hence, we see that in the final state  \eqref{form2measurements} S is not projected to either of the measured quantities, but rather to a linear combination of them. This is yet another signature of the non-ideality of the process. Note also that there are two possible collapse basis, given by $\ketupepp, \ketupemm $ and $\ketupepm, \ketupemp$. Those basis are established by the strength of the interaction with each magnet, as given by \eqref{defketupee}.   

\subsection{Generalizations and dependence on the initial state of the magnet}
While in our results we have assumed  an initial paramagnetic state for the magnets, given by \eqref{RMinitial}, it is easy to see that our considerations naturally apply for other initial states.  First of all, notice that the free energies obtained in \eqref{FreeEnergyTwoApp} depend only on the equilibrium state, and hence are independent of the initial state. Consequently, the equilibrium points of the magnets, which are shown in Fig. \eqref{FreeEnergyTwoApp}, are also independent of it. This implies that the form \eqref{form2measurements} of the final state holds for \emph{any} initial state of the magnets. The dependence on the initials state comes only through the weights $\Ppp,  \Ppm, \Pmp$ and $ \Pmm$. In order to estimate such weights, which are used to obtain Fig. \ref{Lambda}, we have resorted to the equations of motion derived in Appendix \ref{DerivationEqsMotion}. The derivation of such equations of motion depends strongly on the form \eqref{DMM}, which is valid as long as the initial state can be expressed as $\hat{D}_0 = \hat{D}_0 (\hat{m}, \hat{m}')$ (see Appendix \ref{AppendixForm}). That is, we can solve the dynamics of any initial distribution of the magnets that can be expressed as a function of $\hat{m}$ and $\hat{m}'$.

 The technique developed in Sec. \ref{FreeEnergiesSec} (see also Sec. \ref{secFreeEnergyTwoApp})  to construct free energy functions conditioned on the initial state of the spin --from which one obtains \eqref{freeenergyrprm} (for one apparatus) and \eqref{FreeEnergyTwoApp} (for two apparatuses)-- is general and can be applied  to other situations. Indeed, if the metastability of a state is broken when it interacts with another system S, our considerations allow to construct different free energies depending on the possible states of S. Those free energies define the possible final equilibrium states of the metastable state. This technique may find applications not only in quantum measurements, but also in the study of dissipative phase transitions \cite{Cirac} (see also footnote 28 of Ref. \cite{OpusII}).

 \renewcommand{\thesection}{\arabic{section}}
\section{The simultaneous measurement as a generalized quantum measurement}
 \setcounter{equation}{0} \setcounter{figure}{0}
 \renewcommand{\thesection}{\arabic{section}.}

\label{SecPOVM}

In this section we show that this joint measurement process can be well described at an abstract level by a generalized quantum measurement, defined by a Positive-Operator-Valued-Measure (POVM).
This allows us to give a simple operational interpretation of the process.  

Recall that a  POVM is a set of positive operators $\{ F_i \}$, $F_i\geq 0$, which satisfy,
\begin{equation}
\sum_i F_i = \mathbb{I}.
\label{sumFi}
\end{equation}
Our situation will deal with the 2-d case, viz. $ \mathbb{I}=\textrm{diag}(1,1)$.
The probability of the outcome $i$ is then given by 
\begin{equation}
P_i= {\rm Tr}(\rho F_i).
\label{P_i}
\end{equation}
 To determine the post measurement state, the measurement operators need to be expanded as $F_i= M_i^{\dagger} M_i$, 
 and then the post-measurement state for the outcome $i$ takes the form \cite{Nielsen}
\begin{align}
\rho_i = \frac{M_i \rho M_i^{\dagger}}{{\rm Tr}\left(M_i \rho M_i^{\dagger}\right)}.
\label{poststate}
\end{align}

Given these definitions we can express our joint measurement  as a combination of two simple processes at the level of S. Let S be described by a general spin-$\half$ state,
\begin{align}
\rho= \frac{1}{2} \left( \mathbb{I} + \sum_{i=x,y,z} \langle \hat{s}_i (0) \rangle \hat{s}_i \right).
\end{align}
 First, a noisy channel is applied to S, so that $\rho$ turns into
\begin{align}
C(\rho) = \frac{1}{2} \Big(\mathbb{I} +\alpha_x \langle \hat{s}_x (0) \rangle \hat{s}_x + \alpha_z \langle \hat{s}_z (0) \rangle \hat{s}_z   \Big),
\label{rhochannel}
\end{align}
where $\alpha_x, \alpha_z \in [0,1]$ and $\alpha_y=0$ has been assumed already. This corresponds to the loss of information induced by both the dephasing 
effect in \eqref{sxszdecay} and the decoherence induced by the action of the baths.  
After this noisy evolution,  a generalized measurement is applied upon S, given by the following four POVM elements
\begin{align}
F_{\epsilon \epsilon'} = \frac{1}{2} \ketupee \braupee, \hspace{8mm} \epsilon, \epsilon' = \pm.
\label{Feee}
\end{align}
Each $F_{\epsilon \epsilon'}$, corresponds to an outcome of the measurement. To see that the set $\{F_{\epsilon \epsilon'} \}$  defines a POVM, first notice $F_{\epsilon \epsilon'}\geq 0 $. Secondly, by expanding these elements as
\begin{align}
F_{\epsilon \epsilon'} = \frac{1}{2} \left( \Idd + \epsilon' u_x^{(f)} \hat{s}_x    + \epsilon u_z^{(f)} \hat{s}_z   \right)
\label{expansionFee}
\end{align}
we immediately notice that, 
\begin{align}
 F_{++}+F_{--}&= \frac{1}{2}(\ketupepp \braupepp+\ketupemm \braupemm)
= \frac{1}{2}\mathbb{I}, 
\end{align}
and, similarly, 
\begin{align}
 F_{+-}+F_{-+}&= \frac{1}{2}( \ketupepm \braupepm+\ketupemp \braupemp)=\frac{1}{2} \mathbb{I}.
\end{align}
Hence, condition \eqref{sumFi} is satisfied and the set $\{F_{\epsilon \epsilon'}\}$ represent a generalized quantum measurement. We can easily compute the outcome probabilities,  \eqref{P_i}, using the expansion \eqref{expansionFee}, obtaining
\begin{align}
P_{\epsilon \epsilon'} &= \Tr\left(  C(\rho) F_{\epsilon \epsilon'} \right) = 
\nonumber\\
&= \frac{1}{4}\left( 1 + \epsilon' \alpha_x u_x^{(f)}  \langle \hat{s}_x (0) \rangle+ \epsilon \alpha_z u_z^{(f)}  \langle \hat{s}_z (0) \rangle\right) .
\end{align}
This expression is identical to \eqref{Peedeff}  if we identify
 $ \alpha_x=\lambda' / u_x^{(f)}$ and $ \alpha_z=\lambda / u_z^{(f)}$.
 On the other hand, the post-measurement state can be constructed via the operators $M_{\epsilon \epsilon'} = \sqrt{2} F_{\epsilon \epsilon'} $, which satisfy $F_i= M_i^{\dagger} M_i$. Then, by using \eqref{poststate}, we find that the post measurement states are indeed given by $\ketupee$, as in  \eqref{form2measurements}.

The generalized measurement \eqref{Feee} admits a simple interpretation: With probability $1/2$, a projective measurement in the basis spanned by $\ketupepp, \ketupemm$ is applied; and otherwise we apply a projective measurement in the basis of $\ketupepm, \ketupemp$. Hence, from an operational point of view, we can understand the joint measurement as a combination of two projective measurements in which we measure either $\hat{s}_{++}=u_x^{(++)} \hat{s}_x + u_z^{(++)} \hat{s}_z$, or $\hat{s}_{+-}=u_x^{(+-)} \hat{s}_x - u_z^{(+-)} \hat{s}_z$.  Indeed, this combined measurement has four outcomes, with identical probabilities (and corresponding final states) as the dynamical process we consider.
The observables $\hat{s}_{++}$ and $\hat{s}_{+-}$ are a combination of the ``measured" observables $\hat{s}_x$ and $\hat{s}_z$, and the relative weights $u_x^{(f)}$, $u_z^{(f)}$ are determined by the strength of the coupling to each apparatus, as given by \eqref{defuf}. The stronger the coupling to the $z$-component,  the closer $\hat{s}_{+-}$ and $\hat{s}_{-+}$ are to $\hat{s}_z$, and vice versa. 

By expressing the simultaneous measurement as a combination of two (non-commuting) projective measurements in the x-z plane, it easily follows that we can estimate to estimate both $\langle \hat{s}_z (0)\rangle$ and $\langle \hat{s}_x (0)  \rangle$ after many runs of the experiment. Hence the simultaneous measurement is non-ideal but informative: It gives us the average of both  "measured" variables.
It is important to notice, however, that 
in order to employ this POVM approach in practice, it is still necessary to determine $\lambda$ and $\lambda'$ that enter Eq. \eqref{Peedeff}, which
follow from solving the dynamics of the whole measurement, the central theme of the present paper.
In the absence of this knowledge, it is not possible to determine $\langle \hat{s}_x (0) \rangle$ and $\langle \hat{s}_z (0) \rangle$ from the measurement outcomes.
To conclude this section, let us mention an example of a POVM that has been experimentally measured for state discrimination \cite{Mohseni}.

%

 \renewcommand{\thesection}{\arabic{section}}
\section{Conclusion}
 \setcounter{equation}{0} \setcounter{figure}{0}
 \renewcommand{\thesection}{\arabic{section}.}
 
We have studied the possibility of simultaneously measuring two non-commuting spin components using the Curie Weiss model for a quantum measurement, 
developed in \cite{OpusABN,ABNCW}. This model describes a projective measurement of a spin-$1/2$ system as a physical interaction between a system and a magnet, taking the role of the (macroscopic) apparatus. We have worked out the evolution of a spin system simultaneously interacting with two such apparatuses, each of them attempting to measure a different spin component. In order to study the dynamics of this process, we have followed a two-fold approach: In the main text, 
we have derived free energy functions that allows us to infer the form of the final state of the system and the apparatuses, and the main qualitative features of the dynamics involved; and in Appendices 
we have derived rigorously the equations of motion. Combining both methods allows us to gain a qualitative and quantitative understanding of the process. 

We observe an on-going competition between the two apparatuses, each of them trying to obtain information about a different component. This competition appears at different levels, (i) at the beginning of the measurement, when a dephasing effect leads to a partial loss of information of each measured spin component, as shown in (\ref{sxszdecay}), and (ii) during the registration of the measurement, when the  evolution of the pointer state of one apparatus weakens the interaction of the system with the other apparatus (see \eqref{HSMM'}) --this can even prevent the apparatus with the weaker coupling  to achieve a registration at all. 
We have also characterised the minimal interaction system-apparatuses needed for both apparatuses to register results for their respective measurements (see \eqref{h_d} and Fig. \ref{Fi:hc/hd}).

 Even if both apparatuses register a result, the corresponding statistics are imperfect, in the sense that they do not coincide with the ones obtained by separately measuring 
 $\hat{s}_x$ and $\hat{s}_z$ --the two "measured" observables. In other words, one apparatus perturbs the other's measurement via their coupling to the same tested spin, 
 and  the resulting joint measurement is not ideal and can not be described as a simple projective measurement. In order to give an operational interpretation of the obtained 
 statistics, we have constructed in Sec. \ref{SecPOVM} a generalized quantum measurement which provides the same statistics (albeit by possibly different measurement processes).  
 This generalized measurement turns out to be very simple, as it corresponds to a  combination of two projective spin measurements in directions that are specific linear combinations  
 of $\hat{s}_x$ and $\hat{s}_z$. 
It then follows that the resulting information allows us to infer both 
 $\langle \hat{s}_x \rangle$ and $\langle \hat{s}_z \rangle$ with arbitrary precision for sufficiently many runs of the joint experiment.  That is, the measurement is non-ideal but fully informative.

While our results are obtained for a specific  initial (paramagnetic) state for the magnets, the techniques used here can be directly applied to other initial distributions. On the one hand, the equations of motion derived in Appendix \ref{DerivationEqsMotion} can be used for any initial state of the magnets that can be expressed as a function of the magnetizations, $\hat{m}$, $\hat{m}'$. On the other hand, the free energy families  \eqref{FreeEnergyTwoApp} are independent of the initial state and allow one to find the different equilibrium states of the magnets. Hence, the final form    \eqref{form2measurements}  is expected to be generic, depending only on the initial state through the relative weight of each peak. 

A natural extension of the results presented here involves simultaneous measurements involving three apparatuses, in which case a tomographically complete spin measurement is to be expected. It is also interesting to compare our considerations with recent theoretical and experimental results regarding simultaneous measurements \cite{Luis,gills,Shay}. In such works, the "apparatus" is a small quantum system which interacts with the tested system, and is later measured via the standard measurement postulates. In our approach, the full measurement process, including the amplification of the microscopic signal, is treated in a fully quantum mechanical way (the collapse takes place as an effective process due to the many degrees of freedom involved in the apparatuses). Building connections between both approaches, including studies of quantum features of the process \cite{Luis}, would be desirable.


\emph{Acknowledglements}
It is a pleasure to thank Roger Balian for discussion. We also thank Luis Pedro Garc\' ia-Pintos for comments on the manuscript. 
Part of this work was carried out during the master thesis research of M. P.-L. at the University of Amsterdam;
he thanks the members of the Institute for Theoretical Physics for hospitality.
M. P.-L. also acknowledges support from La Caixa Foundation, the CELLEX-ICFO-MPQ Research Fellowships, the Spanish
MINECO (Project No. FIS2013-40627-P and FOQUS
FIS2013-46768-P, Severo Ochoa grant SEV-2015-0522), and the Generalitat
de Catalunya (SGR875).


\myskip{ \appendix

\begin{widetext}
\setcounter{section}{0}
\renewcommand{\thesection}{\roman{section}.}
\section{AppendixA: The Hamiltonian}
\setcounter{equation}{0} \setcounter{figure}{0}
}


\newpage

\appendix

\setcounter{section}{0}
 \renewcommand{\thesection}{\Alph{section}}
  \renewcommand{\thesubsection}{\thesection\arabic{subsection}}
\section{The Hamiltonian}
 \renewcommand{\thesection}{\Alph{section}.}
   \renewcommand{\theequation}{\thesection\arabic{equation}}

\label{AppHamiltonian}


The Hamiltonian of the full system S+A for the single-apparatus case can be split into
\begin{equation}
	\hat{H}_T=\hat{H}_{\rm S}+\hat{H}_{\rm A}+\hat{H}_{\rm SA}
	\label{totalH}
\end{equation}
The internal Hamiltonian of S is neglected, $\hat{H}_{\rm S}=0$, relying on the fact that the measurement happens fast. On the other hand, $\hat{H}_{\rm SA}$ is given by:
\begin{equation}
	\hat{H}_{{\rm SA}}=-g\hat{s}_{z}\sum_{n=1}^{N}\hat{\sigma}_{z}^{\left(  n\right)  }=-Ng\hat{s}_{z}\hat{m}
	\label{HSA}
\end{equation}
where $g>0$ is the strength of the coupling and $\hat{m}$ is the magnetization.  The Hamiltonian of the apparatus, $\hat{H}_{\rm A}$, can be decomposed into,
\begin{equation}
	\hat{H}_{\rm A}=\hat{H}_{{\rm M}}+\hat{H}_{{\rm B}}+\hat{H}_{{\rm M}{\rm B}}
	\label{H_A}
\end{equation}
where $\hat{H}_{\rm M}$ ($\hat{H}_{\rm B}$) is the Hamiltonian of the magnet (bath) and $\hat{H}_{{\rm M}{\rm B}}$ is the coupling between them. $\hat{H}_{\rm M}$, following the Ising model with quartic interactions, is given by:
\begin{equation}
H_{\rm M}=-J_2 N \frac{\hat{m}^2}{2} - J_4 N \frac{\hat{m}^4}{4}.
\end{equation}
Thus the interaction between spins is ferromagnetic ($J>0$), anisotropic (only acts on the z-direction) and couples all the spins $\hat{\sigma}_z^{(i)}$ symmetrically. The quartic interaction ensures metastability.

Each of the $N$ spins $\hat{\sigma}_{a}^{\left(  n\right)}$ of the magnet is coupled to the bath of phonons independently as:
\begin{equation}
	\hat{H}_{{\rm M}{\rm B}}  =\sqrt{\gamma}\sum_{n=1}^{N}\sum_{a=x,y,z}\hat{\sigma}_{a}^{\left(  n\right)  }\hat{B}_{a}^{\left(n\right)  }
	\label{couplingBM}
\end{equation}
where $\hat{B}_{a}^{\left(n\right)} $ are phonon operators given by:
\begin{equation}
		\label{Ba}
		 \hat{B}_{a}^{\left(n\right)  }=\sum_{k}\sqrt{c\left(  \omega_{k}\right)  }\left(  \hat{b}_{k,a}^{\left(  n\right)  }+\hat{b}_{k,a}^{\dagger\left(  n\right)  }\right)
\end{equation}
Then, the diagonalized Hamiltonian of the bath of phonons $\hat{H}_{{\rm M}{\rm B}} $ is:
\begin{equation}
	\hat{H}_{{\rm B}}  =\sum_{n=1}^{N}\sum_{a=x,y,z}\sum_{k}\hbar\omega_{k}\hat{b}_{k,a}^{\dagger\left(  n\right)  }\hat{b}_{k,a}^{\left(  n\right)  }
	\label{H_{B}}
\end{equation}
where $\hat{b}_{k,a}^{\dagger\left(  n\right)  }$ are Debye phonon modes with eigenfrequencies $\omega_k$ (acting on $\hat{\sigma}_{a}^{\left(  n\right)}$). 

The action of B in the dynamics of the S+M is compressed into its autocorrelation function $K(t-t')$.
It is useful to introduce the Fourier transform and its inverse
\begin{eqnarray}
\tilde{K}\left(  \omega\right)
=\int_{-\infty}^{+\infty}{\rm d}t\ e^{-i\omega t}K\left(  t\right),\qquad 
\nonumber\\
K(t)=\frac{1}{2\pi}\int_{-\infty}^{+\infty}{\rm d}\omega \,e^{i\omega t}\tilde K\left(  \omega\right).
\end{eqnarray}
so that $\tilde{K}$ is chosen to have the quasi-Ohmic form \cite{breuer},
\begin{equation}
\tilde{K}\left(
\omega\right)  =\frac{\hbar^{2}}{4}\frac{\omega e^{-\left\vert \omega
\right\vert /\Gamma}}{e^{\beta\hbar\omega}-1}{\rm  .} 
\label{tildeK}
\end{equation}
where the Debye's cutoff $\Gamma$ is the largest frequency of the bath, and it is assumed to be larger than all other frequencies entering the problem.

 \renewcommand{\thesection}{\Alph{section}}
  \renewcommand{\thesubsection}{\thesection\arabic{subsection}}
\section{Dynamics of a measurement of two observables}
 \renewcommand{\thesection}{\Alph{section}.}
   \renewcommand{\theequation}{\thesection\arabic{equation}}
\label{DynamicsTwoApp}

\subsection{The Hamiltonian}
The second apparatus A$'$ is built in close analogy with $A$, in such a way that $A'$ made up of magnet M$'$ and a bath B$'$, with parameters $ J_2', J_4', g', N'...$,  which we assume to have the same order than those of A, and internal Hamiltonian $H_{\rm M'}$ analogue to (\ref{H_A}).  As discussed in the main text, the coupling Hamiltonian between SAA$'$ reads, 
\BEQ
\hat{H}_{\rm SMM'}=-Ng \hat{s}_z \otimes \hat{m} \otimes \mathbb{I}_{M'} -N'g' \hat{s}_x \otimes \mathbb{I}_M \otimes \hat{m}'.
\label{inttwo}
\EEQ

\subsection{Characterization of the state}

When solving the Liouville equation of motion, we use that the state $\hat{D}$ of S+M+M$'$ is a function of $\hat{m}$, $\hat{m}'$, because of  the symmetric properties of the initial paramagnetic state and the Hamiltonian (see \citep{OpusABN}). Besides the characterization given in the main text, a useful characterization for $\hat{D}$ reads,
\BEA
&& \hat D_{SMM'}(m,m',t)=\sum_{m,m'} \frac{1}{G( m)G( m')} \\
 &&\times\left[ P( m,m',t)\frac{\mathbb{I}}{2}+{\bf C}( m,m',t)\cdot\hat{\bf s}\right] \otimes \Pi_m \otimes \Pi_{m'}. \nn
\label{charstatetwonon}
\EEA
where $G$ is the degeneracy of the magnetization. In order to interpret this description, notice that,
\begin{align}
 &P(m,m',t)= \tr \left( \hat{\Pi}_m  \otimes \hat{\Pi}_{m'} \hat{D}_{SMM'}(m,m',t) \right), 
 \nonumber\\
&C_i(m,m',t)= \tr_{\rm S}\left(\hat{s}_i  \otimes \hat{\Pi}_m  \otimes \hat{\Pi}_{m'} \hat{D}_{SMM'}(m,m',t)\right),
\end{align}
Therefore, $P(m,m',t)$ is the joint probability distribution of the magnetization of the apparatuses and 
$C_i(m,m',t)$, with $i=x,y,z$ or $i=x,u,v$; brings information about the correlations between $\hat{s}_i$ and the apparatuses.

Recall that the initial conditions are given by,
\begin{align}
P(m,m',0)&= P_0(m) P_0'(m')
\nonumber\\
C_i(m,m',0)&= \langle s_i \rangle P_0(m) P_0'(m')
\end{align}
where $ \langle s_i \rangle =  \Tr (\rho_S \hat{S}_i)$. Similarly, $P_M(m)$ is the distribution of the initial paramagnetic state, which, for large $N$, is well approximated by a Gaussian distribution, 
\begin{equation}
P_{0}\left(  m\right)  
=\sqrt{\frac{N}{2\pi}}e^{-Nm^2/2}.
\label{Pm0II}
\end{equation}
and $P_0'(m')$ is obtained by replacing $N \leftrightarrow N'$. 

This characterization is explicitly related to the one used in the main text by,
\begin{align}
\Puu^{(u)}&=\frac{P + C_u}{2},
\nonumber\\
\Pdd^{(u)}&=\frac{P- C_u }{2},
\nonumber\\
\Pud^{(u)}&=\frac{\hat{C}_v-i\hat{C}_y}{2},
\nonumber\\
\Pdu^{(u)}&=\frac{\hat{C}_y+i\hat{C}_v}{2}.
\end{align}
where the dependence of the functions over $(m,m',t)$ is implicit, and so will remain for the next computations.

\subsection{ Dephasing}

Let us discuss describe the dephasing process for the two-apparatuses case. This takes place in a short time-scale, where the effect of the baths can be neglected due to $\gamma\ll 1$. The relevant Hamiltonian is then:
\begin{equation}
 \hat{H}=-\frac{\hbar}{2} w(\hat{m},\hat{m}') \hat{s}_{\bf u}+\hat{H}_M(\hat{m})+\hat{H}_{M'}(\hat{m}')
 \label{Hmmm}
\end{equation}
with $\hat{H}_M$, $\hat{H}_{M'}$ being the internal Hamiltonians of the magnets.

Under the Hamiltonian \eqref{Hmmm}, the dynamics can be readily solved and, as shown in the main text, we obtain,
\begin{align}
 P_{\uparrow\downarrow}^{(\bf{u})}(m,m',t)&=e^{i\omega(m,m')t}P(m) P(m') r_{\uparrow \downarrow}^{(\bf{u})} 
 \nonumber\\
  P_{\uparrow\uparrow}^{(\bf{u})}(m,m',t)&=P(m) P(m') r_{\uparrow \uparrow}^{(\bf{u})} 
\end{align}
Or, equivalently in the decomposition \eqref{charstatetwonon},
\begin{align}
&P(t)=P(0) \nonumber\\
&C_u(t)=C_u(0) \nonumber\\
&C_v(t)=C_v(0) \cos(wt) \nonumber\\
&C_y(t)=C_y(0) \sin(wt) 
\label{solution2apwithoutbath}
\end{align}

The expected values of $\hat{s}_i$  can be now computed as,
\begin{align}
\langle \hat{s}_i (t)\rangle=  \sum_{m,m'} C_i(t). 
\end{align}
In the limit, $N,N'\rightarrow \infty$, we can substitute the sum for an integral with the gaussian distribution \eqref{Pm0II} for the initial distribution. Then,  for $Ng=N'g'$, we can analytically solve the different integrals. On the one hand, we obtain,
\begin{align}
\langle \hat{s}_v (t)\rangle= &\langle \hat{s}_v (0)\rangle \int \hspace*{-2mm} \int \hspace{-1mm} dm dm' P_{\rm M}(m) P_{M}(m') \cos\left(wt\right) 
\nn \\ =&\langle \hat{s}_v (0)\rangle\frac{\sqrt{\pi} t}{2 \tau_{\rm d}} e^{-\left(\frac{t}{2\tau_{\rm d}}\right)^2} \\
 \langle \hat{s}_y (t)\rangle =& \langle \hat{s}_y (0)\rangle \int \hspace*{-2mm} \int \hspace{-1mm} dm dm' P_{\rm M}(m) P_{M}(m') \sin\left(wt\right)  
 \nn\\ =& \langle \hat{s}_y (0)\rangle\left(1+e^{-\left(\frac{t}{\tau_{\rm d}}\right)^2} i \sqrt{\pi} \frac{|t|}{\tau_{\rm d}} {\rm erf}\left( i \frac{|t|}{\tau_{\rm d}} \right)\right) 
\end{align}
where erf is the error function and we recall that $\tau_{\rm d} = 1/\sqrt{2N}g$. Both functions decay on a timescale $\tau_{\rm d}$.
  Therefore, we obtain an effective decay due to the rapid oscillating terms in (\ref{solution2apwithoutbath}). 
For $Ng\neq N'g'$, it can be numerically checked that the same mechanism takes place: The oscillatory terms induce a decay of $\langle \hat{s}_v (t)\rangle$ and $\langle \hat{s}_y (t)\rangle$. Let us now turn our attention to the measured components, $x$ and $z$.  For the $x$ direction,
\begin{equation}  
 \langle \hat{s}_x (t)\rangle=\int dm dm' \left[u_x C_u (0) + u_z C_v (0) \sin(wt)\right]
\end{equation}
The time dependent part was argued before to tend to zero in a time scale $\tau_{\rm d}$. Then,
\begin{align}
	\langle \hat{s}_x (t)\rangle&  \xrightarrow{t  \gg \tau_{\rm d}} \int dm dm' (\langle \hat{s}_x (0)\rangle u_x^2 +\langle \hat{s}_z (0)\rangle u_z u_x) P_{\rm M} P_{M'}
	\nn\\ &=\langle \hat{s}_x (0)\rangle \frac{\sqrt{N'}g'}{\sqrt{N}g+\sqrt{N'}g'}, 
\end{align}
Proceeding similarly for the $z$ component,
\begin{eqnarray}
	\langle \hat{s}_z (t)\rangle  \xrightarrow{t  \gg \tau_{\rm d}}  \langle \hat{s}_z (0)\rangle \frac{\sqrt{N}g}{\sqrt{N}g+\sqrt{N'}g'}, 
\end{eqnarray}
Hence we obtain the results \eqref{sxszdecay} announced in the main text. 

Finally, let us briefly discuss the effect of the bath B on the off-diagonal terms. The equilibrium state of SMM' at temperature $1/\beta$ reads
\begin{eqnarray}
\hat{D}_{eq}=\frac{1}{\mathcal{Z}}\left(e^{-\beta(\frac{\hbar}{2}w  \hat{s}_{\bf  u}+\hat{H}_{\rm M}+\hat{H}_{M'})}  \right),
\label{eqstatetwoappp}
\end{eqnarray}
which contains no off-diagonal terms in the $\hat{s}_{\bf u}$ basis. Therefore, if we start in a state out of equilibrium, the bath tends to eliminate the correlators $C_v$, $C_y$, thus increasing the dephasing effect. 


\subsection{Dynamical equations for the registration}
\label{DerivationEqsMotion}
We now proceed to solve the dynamics of the registration, where the bath plays an essential role --for a very detailed derivation we refer the reader to \cite{marti}. Consider thus the quantum state $\hat{\mathcal{D}}(t)$ of the whole system SMM$'$BB$'$. We are interested in $\hat{D}(t)=\tr_{\rm B,B'} \hat{\mathcal{D}}(t)$, which can always be decomposed as \eqref{charstatetwonon} with $i=x,y,z$
Tracing out B and B$'$ from the equation of motion of $\hat{\mathcal{D}}(t)$ yields the formal expression,
\begin{align}
 i\hbar \frac{d \hat{D}}{dt} =& [\hat{H}_{SM}+\hat{H}_{SM'},\hat{D}]+\tr_{\rm B} [\hat{H}_{MB},\tr_{B'} \hat{\mathcal{D}}] \nn\\
 +&\tr_{B'} [\hat{H}_{MB'},\tr_{B} \hat{\mathcal{D}}]
\label{2formalexp}
\end{align}
The second term can be reduced to
\begin{align}
 \tr_{\rm B} [\hat{H}_{MB},\tr_{B'} \hat{\mathcal{D}}]&=i \gamma \sum_{a=x,y,z}\sum_n \int_0^t du\big\{\left[\hat{\sigma}_a^{(n)}(u)\hat{D},\hat{\sigma}_a^{(n)}\right]K(u)
\nonumber\\ &+\left[\hat{\sigma}_a^{(n)},\hat{D}\hat{\sigma}_a^{(n)}(u)\right]K(-u)\big\}.
\label{2noncomeqbath}
\end{align}
with
\begin{equation}
 \hat{\sigma}_a^{(n)}(t)=U_0 \hat{\sigma}_a^{(n)} U^{\dagger}_0, \hspace{10mm} U_0=\exp\left\{-i\frac{t}{\hbar} \hat{H}_0 \right\}
\end{equation}
with $\hat{H}_0=\hat{H}_{SA}+\hat{H}_{SA'}+\hat{H}_{\rm M}+\hat{H}_{M'}$.
Only terms with $\hat{\sigma}_x^{(n)}$ and $\hat{\sigma}_y^{(n)}$ contribute in (\ref{2noncomeqbath}), and it is useful to rewrite them in terms of lowering and raising operators:
\begin{eqnarray}
\hspace{-7mm} \tr_{\rm B} [\hat{H}_{MB},\tr_{B'} \hat{\mathcal{D}}]&=&\sum_n \int_0^t du\Big\{\Big(\Big[\hat{\sigma}_{+}^{(n)}(u)\hat{D},\hat{\sigma}_-^{(n)}\Big] \\
&+&\Big[\hat{\sigma}_-^{(n)}(u)\hat{D},\hat{\sigma}_+^{(n)}\Big]\Big)K(u)+h.c.\Big\}, \nn
\label{expbathsi}
\end{eqnarray}

The computation of $\hat{\sigma}_+^{(n)}(t)$, $\hat{\sigma}_-^{(n)}(t)$ is hindered by the non-commuting terms of $\hat{\hat{H}}_0$. 
Using $\sigma_+ f(\hat{m})=f(\hat{m}+\delta m)$ with $\delta m= 2/N$, we obtain,
\begin{equation}
 \hat{\sigma}_+^{(n)}(t)=\hat{\sigma}_+^{(n)} e^{-i\frac{t}{\hbar} \hat{H}_0(\hat{m}+\delta m)} e^{i\frac{t}{\hbar} \hat{H}_0(\hat{m})}
\label{expsigmaas}
\end{equation}
which can be simplified by using $\exp{i{\bf a}\hat{\bf{s}}}=\cos a +i ({\bf a}\cdot \hat{\bf{s}}/a) \sin a $ and expanding the phase $a$ in powers of $1/N$, 
yielding the leading term 
\begin{eqnarray}
\hspace{-9mm}\hat{\sigma}_+^{(n)}(t)= \hat{\sigma}_+^{(n)} \exp \hspace{-0.6mm} \left\{i\frac{2t}{\hbar}\left(J\hat{m}^3+g u_z {\bf u \cdot \hat{s}}\right)\hspace*{-0.5mm}\right\}.
\label{expevsigmaun}
\end{eqnarray}
We can then find $\hat{\sigma}_-^{(n)}(t)$ by using $\sigma_-=\sigma_+^{\dagger}$. In a similar way we can find the leading term of the 
evolution of the operators $\hat{\sigma}'{}_+^{(n)}(t)$, 
belonging to B$'$:
\begin{eqnarray}
\hspace*{-4mm}\hat{\sigma}'{}_+^{(n)}(t)=\exp\left\{i\frac{2t}{\hbar}\left(J'\hat{m}'{}^3+g' u_x {\bf u \cdot \hat{s}}\right)\right\}.
\label{expevsigmaunII}
\end{eqnarray}

Now we can insert (\ref{expevsigmaun}) and \eqref{expevsigmaunII} into (\ref{expbathsi}) to drop the time dependence of the operators $\hat{\sigma}_+^{(n)}(u)$. The time dependence is then be found in integrals of the form 
\begin{align}
 \tilde{K}_{t>}\left(  \omega\right) &=\int_{0}
^{t}{\rm d} u e^{-i\omega u}K\left(  u\right) \nn\\
&=\frac{1}{2\pi i}\int
_{-\infty}^{+\infty}{\rm d}\omega^{\prime}\tilde{K}\left(  \omega^{\prime}\right)
\frac{e^{i\left(  \omega^{\prime}-\omega\right)  t}-1}{\omega^{\prime}-\omega}
 {\rm  ,} 
\label{kt>}
\end{align}
and $\tilde{K}_{t<} \left(  \omega\right)  =\left[  \tilde{K}_{t>}\left(
\omega\right)  \right]  ^{\ast}$. It is useful to define:
\begin{align}
&\tilde{K}_t(w)=\tilde{K}_{t>}(w)+\tilde{K}_{t<}(w) \\
&\tilde{K}_t'(w)=i(\tilde{K}_{t>}(w)-\tilde{K}_{t<}(w))
\label{KandK'def}
\end{align}
We can then insert (\ref{charstatetwonon}) into (\ref{expbathsi}) and make use of the functionality dependence on $m, m'$ in order to get scalar equations for $P$, $C_x$, $C_y$ and $C_z$ (this dependence is justified in Sec. \ref{AppendixForm}). Defining 
\begin{align}
&\Delta_+ f(m)= f(m+\delta m)- f(m)
\nonumber\\
&\Delta_- f(m)=f(m-\delta m)-f(m),
\end{align}
we obtain the following equations of motion,
\begin{widetext}
\begin{align}
& \frac{\partial P}{\partial t}=\frac{\gamma N}{2 \hbar^2} \left[ \Delta_+ \left\{\alpha_+ P+\beta_+ u_i C_i \right\}+
\Delta_- \left\{\alpha_- P +\beta_- u_i C_i \right\}
 \right]+ \mathrm{B'\hspace{1mm} terms}  
\nonumber\\
\nonumber\\
&\frac{\partial C_x}{\partial t}-w u_z C_y 
=\frac{\gamma N}{2 \hbar^2} \left[ \Delta_+ \left\{\beta_+ u_x P+\alpha_+  C_x + \kappa_+ C_y u_z\right\}
+\Delta_- \left\{\beta_- u_x P+\alpha_-  C_x + \kappa_- C_y u_z\right\}\right] + \mathrm{B'\hspace{1mm} terms}
\nonumber\\
\nonumber\\
&\frac{\partial C_y}{\partial t}+\hspace{-0.5mm} w (u_z C_x-u_x C_z) \hspace{-0.5mm}
= \hspace{-0.5mm}\frac{\gamma N}{2 \hbar^2} \left[ \Delta_+\hspace{-0.5mm} \left\{\alpha_+  C_y - \kappa_+(C_zu_x+u_z C_x)\right\}
+\Delta_-\hspace{-0.5mm}\left\{\alpha_-  C_y - \kappa_-(C_zu_x+u_z C_x)\right\}
 \right]\hspace{-0.5mm} +\hspace{-0.5mm}\mathrm{B'\hspace{0.5mm} terms}
\nonumber\\
\nonumber\\
&\frac{\partial C_z}{\partial t}+w u_x C_y 
=\frac{\gamma N}{2 \hbar^2} \left[ \Delta_+ \left\{\beta_+ u_z P+\alpha_+  C_z + \kappa_+ C_y u_x\right\}
+\Delta_- \left\{\beta_- u_z P+\alpha_-  C_z + \kappa_- C_y u_x\right\}\right] + \mathrm{B'\hspace{1mm} terms}
\label{eqmotio2apparatus}
\end{align}
\end{widetext}
where we defined:
\begin{eqnarray}
\alpha_+&=&(1+m)\left[\tilde{K}_t(2\omega_+)+\tilde{K}_t(2\omega_-)\right] ,\nonumber\\
\alpha_-&=&(1-m)\left[\tilde{K}_t(-2\omega_+)+\tilde{K}_t(-2\omega_-)\right] ,\nonumber\\
\beta_+&=&(1+m)\left[\tilde{K}_t(2\omega_+)-\tilde{K}_t(2\omega_-)\right] ,\nonumber\\
\beta_-&=&(1-m)\left[\tilde{K}_t(-2\omega_+)-\tilde{K}_t(-2\omega_-)\right] ,\nonumber\\
\kappa_+&=&(1+m)\left[\tilde{K}_t'(2\omega_-)-\tilde{K}_t'(2\omega_+)\right] ,\nonumber\\
\kappa_-&=&(1-m)\left[\tilde{K}_t'(-2\omega_-)-\tilde{K}_t'(-2\omega_+)\right],
\label{definitionsabk}
\end{eqnarray}
and
\begin{eqnarray}
\hbar \omega_{\pm}=J_2 m+J_4 m^3 \pm g u_z.
\label{defomeganotnorm}
\end{eqnarray}
The $B'$ terms, which are the terms arising from the second bath, have the same form as the terms arising from the first bath but replacing $N\rightarrow N'$ and $\gamma \rightarrow \gamma'$ in the terms of the equations of motion,  taking the differences $\Delta_{\pm}$ over $m'$, and replacing $J\rightarrow J'$, $m\rightarrow m'$, $g\rightarrow g'$ and $N\rightarrow N'$ in the definitions (\ref{definitionsabk}) and (\ref{defomeganotnorm}). In particular,
\begin{eqnarray}
\hbar \omega_{\pm}=J'_2 m'+J'_4 m'^3 \pm g' u_x.
\label{defomeganotnorm}
\end{eqnarray}

Notice from expression \eqref{defomeganotnorm} that the original strength of the coupling $g, g'$ of S to the apparatuses is effectively weakened to $g u_z, g' u_x$. Recall that $u_x,u_z$ are given by
\begin{align}
& u_x=\frac{N'g'm'}{\sqrt{(Ngm)^2+(N'g'm')^2}} , \nn\\
& u_z=\frac{Ngm}{\sqrt{(Ngm)^2+(N'g'm')^2}} ,
\end{align}
where we note that one measurement hinders the other. For example, a measurement in the apparatus A, which implies an increment of $|m|$, will increase $g u_z$ while decreasing $g' u_x$.


\subsubsection{Markovian regime}
In the registration process, the bath drives the magnet to a stable state. Such a process takes place in a timescale $\tau_{\gamma}=\hbar/\gamma J$, $\tau_{\gamma'}=\hbar/\gamma' J'$ for A and A' respectively. Since the interaction with the baths is weak, $\gamma,\gamma' \ll 1$, the Markovian condition ( $\tau_{\gamma} \gg \hbar/2 \pi T$) is satisfied. In such a regime the integrals $\tilde{K}_t (w) \rightarrow \tilde{K}(w)$ and $\tilde{K}'_t (w)$ lose their time dependence becoming:
\begin{equation}
  \tilde{K}(w)=\frac{\hbar^2}{4} \frac{w e^{-\frac{|w|}{\Gamma}}}{e^{\beta \hbar w}-1}
\end{equation}
and $ \tilde{K}'_t (w) \rightarrow \tilde{K'}(w)$ with
\begin{eqnarray}
 \tilde{K}'(w)=-\frac{ \hbar^2}{2}\left[\frac{1}{\beta \hbar} \sum_{n=1}^{\infty} e^{-\frac{\Omega_n}{\Gamma}} \frac{\Omega_n}{w^2+\Omega_n^2}+\frac{\tilde{\Gamma}}{4} \right] ,
\end{eqnarray}
where $\Omega_n=\frac{2\pi n}{\hbar \beta}$. Notice that the second term of $\tilde{K}'(w)$ is a constant and it depends linearly on $\tilde{\Gamma}$, so that it is bigger than any other term encountered so far. Nevertheless, since $\tilde{K}'(w)$ appears in the equations of motion in the form $\tilde{K}'(a)-\tilde{K}'(b)$, as we can see from (\ref{definitionsabk}), this constant terms drop out. 

The equations of motion (\ref{eqmotio2apparatus}) involve two different time scales: $\tau_{g}=\hbar/g$ and $\tau_{\gamma}=\hbar/\gamma J$  corresponding to the couplings SM  and MB,  respectively -we could also have chosen the time scales corresponding to the second apparatus A$'$, but those are assumed to be of the same order. In the studied model it is satisfied that $\gamma \ll g/J$, so that $\tau_{g}\ll \tau_{\gamma}$. Let us now consider the equations of motion for $P, C_u, C_v$ and $C_y$, which are obtained by taking appropriate linear combinations of the equations of motion (\ref{eqmotio2apparatus}). From such equations, one can see that $P$ and $C_u$ evolve slowly, only under the effect of the baths, whereas $C_y$ and $C_v$ evolve fastly under the effect of the coupling SMM$'$. Hence, effectively the slow variables $P, C_u$ only depend on the average of the fast variables $C_y$ and $C_v$ over the short time scale.

Disregarding the effect of the bath, the evolution in time of $C_v$ and $C_y$ was found in (\ref{solution2apwithoutbath}). The solution has an oscillatory nature with frequency $w(m,m')$. For typical values of $m$ and $m'$ of order $1/\sqrt{N}$,$1/\sqrt{N'}$, we have that $w \sim \mathcal{O}(1/\sqrt{N},1/\sqrt{N'})$. Then averaging the solutions over the short time scale $\tau_g$ yields
\begin{eqnarray}
\hspace{-5mm} \langle C_y(t) \rangle_{\tau_g}=C_y(0) 
  \frac{\sin{w\tau_g}}{w\tau_g} \sim \mathcal{O}\left(\frac{1}{\sqrt{N}},\frac{1}{\sqrt{N'}}\right)
\nonumber\\
\hspace{-5mm} \langle C_v(t) \rangle_{\tau_g}= C_v(0) \frac{\cos{w\tau_g}}{w\tau_g} \sim \mathcal{O}\left(\frac{1}{\sqrt{N}},\frac{1}{\sqrt{N'}}\right).
\end{eqnarray}
Since $N,N' \gg 1$, the evolution of the slow variables $P$ and $C_u$ is independent of the fast variables. 
Therefore in the Markovian regime $P$ and $C_u$ evolve according to the much simpler dynamics,
\begin{widetext}
\begin{align}
 \frac{\partial P}{\partial t}=\frac{\gamma N}{2 \hbar^2} \left[ \Delta_+ \left\{\alpha_+ P+\beta_+ C_u \right\}+
\Delta_- \left\{\alpha_- P +\beta_- C_u \right\}
 \right]+ \mathrm{B'\hspace{1mm} terms}  
\nonumber\\
\nonumber\\
\frac{\partial C_u}{\partial t}=\frac{\gamma N}{2 \hbar^2} \left[ u_x \Delta_+ \left\{u_x(\alpha_+ C_u+\beta_+ P) \right\}+u_z \Delta_+ \left\{u_z(\alpha_+ C_u+\beta_+ P) \right\}
\right.
\nonumber\\
\left. \hspace{6mm}
+u_x \Delta_- \left\{u_x(\alpha_- C_u+\beta_- P) \right\}+u_z \Delta_- \left\{u_z(\alpha_- C_u+\beta_- P) \right\}
 \right]+ \mathrm{B'\hspace{1mm} terms} .
\label{secondeqmotiontwoobservables}
\end{align}
\end{widetext}
These coupled equations can be easily solved by numerical methods for large $N$ and $N'$  --in Figures \ref{Fi:plotPIIdfas} and \ref{Fi:plotPIIdfasII},   $N=N'=161$. The figures and numerical estimates of the main text are based upon the equations \eqref{secondeqmotiontwoobservables}.

%

\subsubsection{A Fokker-Plank equation for the process}
In this section we apply some further simplifications to the equations of motion in order to get a Fokker-Plank like equation of motion. This allows us, in a simplified scenario, to solve analytically the equations of motion. We also discuss limitations of this derivation. 

Consider the equation of motion for $C_u$, given in (\ref{secondeqmotiontwoobservables}). First notice that the differences $\Delta_\pm$ satisfy the relation,
\begin{align}
 \Delta_\pm [f(m)g(m)]=&\left[\Delta_\pm f(m)\right] g(m) + f(m) \left[\Delta_\pm g(m)\right] 
\nn\\ +& [\Delta_\pm f(m)] [\Delta_\pm g(m)].
\label{relatioDeltapm}
\end{align}
Let us then apply (\ref{relatioDeltapm}) repeatedly to the right hand side of (\ref{secondeqmotiontwoobservables}). The terms containing $\Delta_+$ yield
\begin{widetext}
\begin{align}
&u_x  \Delta_+ \left\{u_x(\alpha_+ C_u+\beta_+ P) \right\}+u_z \Delta_+ \left\{u_z(\alpha_+ C_u+\beta_+ P) \right\}  \nonumber\\
&=\Delta_+ \left\{\beta P+\alpha C_u \right\}-
(\alpha_+ C_u+\beta_+ P+\Delta_+\{\alpha_+ C_u+\beta_+ P\}) 
 (u_x\Delta_+u_x+u_z\Delta_+u_z+(\Delta_+u_x)^2+(\Delta_+u_z)^2)
\label{eqmotionderivinglambda}
\end{align}
\end{widetext}
and similarly for the other terms. If we expand $\Delta_+ u_i$ in powers of $\delta m=2/N$,
\begin{align}
 &\Delta_+ u_z=\frac{u_x^2 u_z}{m}\frac{2}{N}-\frac{6u_x^2u_z^3}{m^2}\frac{1}{N^2}+O(1/N^3) \nonumber\\
 &\Delta_+ u_x=-\frac{u_z^2 u_x}{m}\frac{2}{N}+\frac{4}{m^2}(u_xu_z^4-u_x^3u_z^2)\frac{1}{N^2}+O(1/N^3),
\end{align}
expression (\ref{eqmotionderivinglambda}) becomes,
\begin{eqnarray}
 &&\frac{\gamma N}{2 \hbar^2} \left[ \Delta_+ \left\{\beta_+ P+\alpha_+ C_u \right\}  \right. \\
&& \left. -\left(\alpha_+ C_u+\beta_+ P+\Delta_+\{\alpha_+ C_u+\beta_+ P\}\right)\frac{2u_x^2 u_z^2}{m^2}\frac{1}{N^2}\right] .\nn
 \label{derivationFPEq}
\end{eqnarray}
If we assume that $P$ and $C_u$ are exponential distributions of the type $e^{-N A}$ (recall the initial conditions (\ref{Pm0II})), then $\Delta_+ P $ is of the same order 
in $N$ as $P$. For typical values of $m,m' \sim \mathcal{O} ( 1/\sqrt{N},1/\sqrt{N'})$, the second term of the previous expression can thus be neglected for large N. 
(The validity of this simplification will be discussed in detail later.)
 Proceeding in the same way as for the other terms in the equations of motion (\ref{secondeqmotiontwoobservables}), we reach  
\begin{align}
  \frac{\partial P}{\partial t}&=\frac{\gamma N}{2 \hbar^2} \left[ \Delta_+ \left\{\alpha_+ P+\beta_+ C_u \right\}+
\Delta_- \left\{\alpha_- P +\beta_- C_u \right\} \right] \nn\\
 &+ \mathrm{B'\hspace{1mm} terms}+\mathcal{O}\left(\frac{1}{m^2N^2},\frac{1}{m'^2N'^2} \right) 
\nonumber\\
\nonumber\\
\frac{\partial C_u}{\partial t}&=\frac{\gamma N}{2 \hbar^2} \left[ \Delta_+ \left\{\alpha_+ C_u+\beta_+ P \right\}+
\Delta_- \left\{\alpha_- C_u +\beta_- P \right\}
 \right] \nn\\ &+ \mathrm{B'\hspace{1mm} terms}+\mathcal{O}\left(\frac{1}{m^2N^2},\frac{1}{m'^2N'^2} \right)
\end{align}
Recalling the definitions,
\begin{eqnarray} 
 P_{\uparrow \uparrow}^{(\bf u)}=\frac{P+C_u}{2}, \qquad 
P_{\downarrow \downarrow}^{(\bf u)}=\frac{P-C_u}{2},
\label{definitionP+P-}
\end{eqnarray} 
we obtain two decoupled equations of motion (as we will discuss later, the corrections $\mathcal{O}\left(1/m^2N^2,1/m'^2N'^2 \right)$ couple $ P_{\uparrow \uparrow}^{(\bf u)}$ and $P_{\downarrow \downarrow}^{(\bf u)}$):

\begin{widetext}
\begin{eqnarray}
   \frac{\partial  P_{\uparrow \uparrow}^{(\bf u)}}{\partial t}&=\frac{\gamma N}{ \hbar^2} \left[ \Delta_+ \left\{ P_{\uparrow \uparrow}^{(\bf u)} (1+m) K_t(2w_+)\right\}+
\Delta_- \left\{ P_{\uparrow \uparrow}^{(\bf u)} (1-m) K_t(-2w_+)\right\}
 \right] + \mathrm{B'\hspace{1mm} terms}+\mathcal{O}\left(\frac{1}{m^2N^2},\frac{1}{m'^2N'^2} \right) 
\nonumber\\
\nonumber\\
  \frac{\partial P_{\downarrow \downarrow}^{(\bf u)}}{\partial t}&=\frac{\gamma N}{ \hbar^2} \left[ \Delta_+ \left\{P_{\downarrow \downarrow}^{(\bf u)} (1+m) K_t(2w_-)\right\}+
\Delta_- \left\{P_{\downarrow \downarrow}^{(\bf u)} (1-m) K_t(-2w_-)\right\}
 \right]+ \mathrm{B'\hspace{1mm} terms}
 +\mathcal{O}\left("," \right) .
\end{eqnarray}
\end{widetext}

Now we bring these equations to the continuum limit following \cite{OpusABN}, where such a derivation is made in detail for the case of one apparatus. 
First, we note that, in the continuum limit,  the differences $ \Delta_{\pm}$ are related to derivatives by
\begin{equation}
 \Delta_{\pm}f(m)=f(m \pm \delta m)-f(m)=\sum_{k=-\infty}^{\infty} \frac{\delta m^k}{k!} \frac{\partial f(m)}{\partial m} ,
\label{expansionDpm}
\end{equation}
then, using $N\gg1$ and recalling $\delta m=\frac{2}{N}$, we can keep only the first terms of the expansion. 
By carefully keeping only the dominant terms in $N$, and assuming that $ P_{\uparrow \uparrow}^{(\bf u)}$ and $P_{\downarrow \downarrow}^{(\bf u)}$ have a gaussian-like shape during the dynamics, we finally find  (see \cite{OpusABN} and \cite{marti} for a detailed discussion),
\begin{align}
  \frac{\partial P_{ii}^{(\bf u)}}{\partial t}&= \frac{\partial}{\partial m}[-v_{i} P_{ii}^{(\bf u)}]+\frac{1}{N}\frac{\partial^2}{\partial m^2}[w_{i} P_{ii}^{(\bf u)}] \nn\\
  &+ \frac{\partial}{\partial m'}[-v^{'}_{i} P_{ii}^{(\bf u)}]+\frac{1}{N'}\frac{\partial^2}{\partial m^{'2}}[w^{'}_{i} P_{ii}^{(\bf u)}]
  \nonumber\\
  &+\mathcal{O}\left(\frac{1}{N},\frac{1}{N'} \right)+\mathcal{O}\left(\frac{1}{m^2N^2},\frac{1}{m'^2N'^2}\right),
\label{FPequationTwoApparatusess}
\end{align}
with $i=\{ \uparrow ,  \downarrow\}$ and
\begin{eqnarray}
 v_{i}&=&\gamma\omega_{i}  (1-m\coth\beta\hbar\omega_{i}) , \nonumber\\
 w_{i}&= & \gamma\omega_{i}  (\coth\beta\hbar\omega_{i}-m) , \nonumber\\
 v^{'}_{i}&=& \gamma\omega^{'}_{i}  (1-m^{'}\coth\beta\hbar\omega^{'}_{i}) , \nonumber\\
 w^{'}_{i}&= & \gamma\omega^{'}_{i}  (\coth\beta\hbar\omega^{'}_{i}-m^{'}) ,
\label{expressionsvwtwo}
\end{eqnarray}
which involve the frequencies
\begin{align}
 \hbar \omega_{\uparrow}&=Jm^3 + gu_z,   \qquad
& \hbar \omega_{\downarrow}=Jm^3 - gu_z,   
 \nonumber\\
\hbar \omega^{'}_{\uparrow}&=Jm^{'3}+ gu_x , \qquad
 & \hbar \omega^{'}_{\downarrow}=Jm^{'3}- gu_x.
\end{align}

\subsubsection{Interpretation of the equations of motion}
The Fokker-Planck equations of motion \eqref{FPequationTwoApparatusess} allow for a simple interpretation:  the functions ${\bf v_i}=(v_{i},v^{'}_{i})$ correspond to a vector velocity of the distribution $P_{i}^{(\bf u)}(m,m',t)$ in the vector space $(m,m')$, whereas the functions ${\bf w}_i=(w_{i},w^{'}_{i})$ are dispersion terms  \cite{OpusABN,marti}. We can see ${\bf v_{\uparrow}}$ and ${\bf v_{\downarrow}}$ plotted in Fig. \ref{Vforpositivenegative}.
Observe how they have completely opposite behaviours: Whereas ${\bf v_{\uparrow}}$ tends to move the distribution to the corners $(\pm 1, \pm 1)$, ${\bf v_{\uparrow}}$ tends to move the distribution to the center. This is in good agreement with the dynamics observed by numerically solving the eqs. \eqref{secondeqmotiontwoobservables}, see in particular Figs. \ref{Fi:plotPIIdfas} and \ref{Fi:plotPIIdfasII}. Indeed, we first observe how $P(m,m',t)$ splits into two distributions,  $ P_{\uparrow \uparrow}^{(\bf u)}$ and $P_{\downarrow \downarrow}^{(\bf u)}$. The $ P_{\uparrow \uparrow}^{(\bf u)}$ , which is driven by  ${\bf v_{\uparrow}}$, moves to the corners of $(m,m')$ whereas the other $P_{\downarrow \downarrow}^{(\bf u)}$ moves to the center, to late disappear. In the next sections we discuss such a disappearance, which cannot be described by eq. \eqref{FPequationTwoApparatusess}, as more terms need to be taken into account in the approximation.

As a final remark, however, let us also note the strong parallelism between the field velocities ${\bf v_i}$ and the free energy functions $F_{\pm}(m,m')$ used in the main text. Indeed, both of them predict the same equilibrium points for the distributions $ P_{\uparrow \uparrow}^{(\bf u)}$ and $P_{\downarrow \downarrow}^{(\bf u)}$.

\subsubsection{Discussing a simplified scenario}
In this subsection we discuss the equations of motion in a simplified scenario, in order to describe the disappearance of $P_{\downarrow \downarrow}^{(\bf u)}$. First of all, we assume that the two apparatuses are identical: $N'=N, \gamma'=\gamma, J'=J, g'=g$. In this case, $u_x$ and $u_z$ simplify to 
\begin{eqnarray}
u_z=\frac{m}{\sqrt{m^2+m'^2}} \nonumber\\
u_x=\frac{m'}{\sqrt{m^2+m'^2}}
\end{eqnarray}
It is useful to make the change of variables:
\begin{eqnarray}
m= r \cos{\theta} \nonumber\\
m'=r \sin{\theta}
\end{eqnarray}
so that $u_x=\sin{\theta}$ and $u_z=\cos{\theta}$. Furthermore, let us work in polar coordinates,
\begin{eqnarray}
v_{i}^{(r)}=v_{i} \cos{\theta}+v_{i}^{'} \sin{\theta} \nonumber\\
v_{i}^{(\theta)}=v_{i}^{'} \cos{\theta}-v_{i} \sin{\theta}
\end{eqnarray}
with $i=\{\uparrow, \downarrow \}$, and 
where $v_{r}$ and $v_{\theta}$ are the radial and the angular velocity respectively.

Secondly, we assume that the field velocity has radial symmetry 
\begin{eqnarray}
 v_{i}=v_{i}^{(r)}(r) \cos{\theta} \nonumber\\
v_{i}'=v_{i}^{(r)}(r) \sin{\theta}
\end{eqnarray}
This condition is satisfied for small times, when ($m$,$m'$) are close to zero. Indeed, if $m,m' \rightarrow 0$, only the interaction with $g$ contributes so that
\begin{eqnarray}
 v_{i}\rightarrow \pm \frac{\gamma}{\hbar} g \cos{\theta} \nonumber\\
v^{'}_{i}\rightarrow \pm \frac{\gamma}{\hbar} g \sin{\theta} 
\end{eqnarray}
which has a radial symmetry.
 Thirdly, we shall assume constant dispersion functions $w_{i}$ and $w'_{i}$. This is motivated by noting that $w,w'$ only change slowly with $m,m'$ and thus, for small times, it suffices to assume them to be constant for the present discussion.   Then, for small times and $\beta \gg 1$, we can take,
\begin{equation}
 w_{i},w^{'}_{i}\rightarrow w=\frac{\gamma}{\beta \hbar}.
\label{simplficationw}
\end{equation}
Since the two apparatuses are identical, we have that $w=w'$. Summarizing, the present discussion approximately holds for small times and two identical apparatuses. 
Using such approximations and 
\begin{eqnarray}
 \frac{\partial}{\partial m}=\cos{\theta} \frac{\partial}{\partial r}-\frac{\sin{\theta}}{r}\frac{\partial}{\partial \theta}  \nonumber\\
 \frac{\partial}{\partial m'}=\sin{\theta} \frac{\partial}{\partial r}+\frac{\cos{\theta}}{r}\frac{\partial}{\partial \theta} \nonumber\\
\bigtriangledown^2 = \frac{1}{r} \frac{\partial}{\partial r}+\frac{\partial^2}{\partial r^2}+\frac{1}{r^2}\frac{\partial^2}{\partial \theta^2}
\end{eqnarray}
the Fokker-Plank equation (\ref{FPequationTwoApparatusess}) becomes,
\BEA
 \frac{\partial P_{ii}^{(\bf u)}}{\partial t}&=&- \frac{1}{r}\frac{\partial}{\partial r}\left(v_{i}^{(r)} r P_{ii}^{(\bf u)} \right) \\
& +&\frac{w}{N}\left[\frac{1}{r} \frac{\partial}{\partial r}+\frac{\partial^2}{\partial r^2} + \frac{1}{r^2} \frac{\partial^2}{\partial \theta^2} \right]P_{ii}^{(\bf u)} {\rm .} \nn
\EEA
with $i=\{\uparrow, \downarrow \}$.

\begin{widetext}

\begin{figure}
\vspace{5mm}
  \centering
  \includegraphics[scale=0.37]{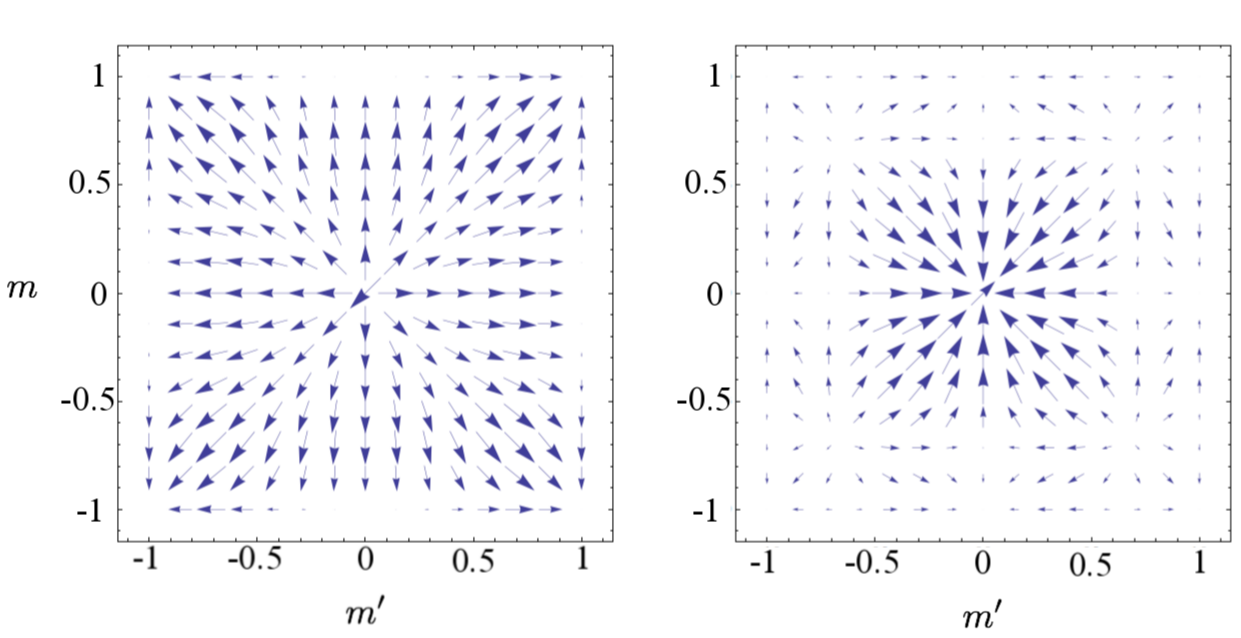}
\caption{Plot of the field velocity of the Fokker-Plank equation associated with $P_{\uparrow \uparrow}^{(\bf u)}$ (left) and $P_{\downarrow \downarrow}^{(\bf u)}$ (right). 
The parameters are chosen as: $T = 0.2J, g = T$ and we take the two appartuses to be identical. The plots clearly show how  $P_{\uparrow \uparrow}^{(\bf u)}$  whereas 
$P_{\downarrow \downarrow}^{(\bf u)}$ tends to the center.}
\label{Vforpositivenegative}
\end{figure}

\end{widetext}

The initial conditions of  $P_{ii}^{(\bf u)}$ is given by,
\begin{align}
 P_{\uparrow \uparrow }^{(\bf u)}=\frac{1}{2}(1+ \langle \hat{s}_z(0) \rangle \cos{\theta} + \langle \hat{s}_x(0) \sin{\theta} \rangle) \frac{1}{2\pi}P_0(r) ,
 \nonumber\\
  P_{\downarrow \downarrow}^{(\bf u)}=\frac{1}{2}(1- \langle \hat{s}_z(0) \rangle \cos{\theta} - \langle \hat{s}_x(0) \sin{\theta} \rangle) \frac{1}{2\pi}P_0(r) ,
\end{align}
with
\begin{equation}
P_0(r)=N \exp\left\{ -\frac{N}{2} r^2 \right\}.
\end{equation}
This suggests the ansatz,
\begin{align}
 P_{\uparrow \uparrow}^{(\bf u)}(r,\theta)=R_{\uparrow \uparrow}(r) + X_{\uparrow \uparrow}(r)\sin{\theta} + Z_{\uparrow \uparrow}(r)\cos{\theta},
 \nonumber\\
  P_{\downarrow \downarrow}^{(\bf u)}(r,\theta)=R_{\downarrow \downarrow}(r) - X_{\downarrow \downarrow}(r)\sin{\theta} - Z_{\downarrow \downarrow}(r)\cos{\theta},
\end{align}
which leads to the independent equations
\begin{align}
& \frac{\partial R_{ii}}{\partial t}=-\frac{1}{r}\frac{\partial}{\partial r}(v^{(r)}_i r R_{ii} ) +\frac{w}{N}\left[\frac{1}{r} \frac{\partial}{\partial r}+\frac{\partial^2}{\partial r^2} \right]R_{ii}, 
\nonumber\\
&\frac{\partial X_{ii}}{\partial t}=-\frac{1}{r}\frac{\partial}{\partial r}(v^{(r)}_i r X_{ii} ) +\frac{w}{N}\left[\frac{1}{r} \frac{\partial}{\partial r}+\frac{\partial^2}{\partial r^2} - \frac{1}{r^2} \right]X_{ii}, 
\nonumber\\
&\frac{\partial Z_{ii}}{\partial t}=-\frac{1}{r}\frac{\partial}{\partial r}(v^{(r)}_i r Z_{ii} ) +\frac{w}{N}\left[\frac{1}{r} \frac{\partial}{\partial r}+\frac{\partial^2}{\partial r^2} - \frac{1}{r^2} \right]Z_{ii}, 
\end{align}
with  $i= \{\uparrow, \downarrow \}$. On the other hand, the weight of a radial distribution $P(r)$ is found by,
\begin{equation}
 \int_0^{\infty}r P(r)\hspace{1mm} dr
\end{equation}
and notice that at $t=0$,
\BEA
 \int_0^{\infty}r P_{ii}^{(\bf u)}\hspace{1mm} dr&=& \int_0^{\infty}r R_{ii}\hspace{1mm} dr=\int_0^{\infty}r X_{ii}\hspace{1mm} dr\nn\\
 &=&\int_0^{\infty}r Z_{ii}\hspace{1mm} dr=1, \hspace{5mm} i=x,z. 
\EEA
From the equations of motion we find that the evolution of such weights in time are given by,
\begin{align}
&\frac{\partial}{\partial t}\int_0^{\infty}r  R_{ii} \hspace{1mm} dr=0
\nonumber\\
&\frac{\partial}{\partial t}\int_0^{\infty}r  X_{ii} \hspace{1mm} dr=-\frac{w}{N}\int_0^{\infty} \frac{1}{r}  X_{ii} \hspace{1mm} dr,
\nonumber\\
&\frac{\partial}{\partial t}\int_0^{\infty}r  Z_{ii} \hspace{1mm} dr=-\frac{w}{N}\int_0^{\infty} \frac{1}{r}  Z_{ii} \hspace{1mm} dr, 
\label{ratelostmemory}
\end{align}
Strictly speaking the limits of the integral range from $2/N$ up to $1$, therefore the the right hand does not diverge. This result shows that $X_{ii}, Z_{ii}$ decrease and therefore the distributions $P_{\pm}$ tend to become symmetric (their dependence on $\theta$ is progressively lost). Therefore the distributions progressively lose memory of the initial conditions $\langle \hat{s}(0) \rangle$. Such loss happens at different rates for $ P_{\uparrow \uparrow}^{(\bf u)}$ and $P_{\downarrow \downarrow}^{(\bf u)}$. Indeed, $ P_{\uparrow \uparrow}^{(\bf u)}$ rapidly flows out from the center and therefore the rate (\ref{ratelostmemory}) deacreses. Furthermore, $w$ tends to zero as $m\rightarrow 1$, so that the rate tends to zero. On the other hand, $P_{\downarrow \downarrow}^{(\bf u)}$ tends to be more peaked at the center, and therefore the rate (\ref{ratelostmemory}) increases and $P_{\downarrow \downarrow}^{(\bf u)}$ rapidly becomes symmetric.

Numerical simulations agree with the found results. They show how $P_{\downarrow \downarrow}^{(\bf u)}$ rapidly becomes symmetric whereas the final distribution of $ P_{\uparrow \uparrow}^{(\bf u)}$ is not. Indeed, the Figs. \ref{Fi:plotPIIdfas} and \ref{Fi:plotPIIdfasII} of $P= P_{\uparrow \uparrow}^{(\bf u)} + P_{\downarrow \downarrow}^{(\bf u)}$ show that the final distribution keeps memory only of the initial conditions of $ P_{\uparrow \uparrow}^{(\bf u)}$. 

Regarding $P_{\downarrow \downarrow}^{(\bf u)}$, we can find the equilibrium distribution by setting $\frac{\partial P^{-}_o}{\partial t}=0$. This leads to:
\begin{equation}
 P_{\downarrow \downarrow}^{(\bf u)}(t\rightarrow \infty)=\frac{(Ng\beta)^2}{2\pi}e^{-Ng\beta r} .
\label{equilibriumdistributionP-} 
\end{equation}
Notice that for such a distribution 
\begin{equation}
 \langle r \rangle=\frac{2}{Ng\beta} .
\end{equation}
Therefore, $m,m'$ take typical values of order $\mathcal{O}(1/N,1/N')$. In such a case, the corrections in (\ref{FPequationTwoApparatusess}) can no longer be neglected. In the following section we discuss such corrections.

\subsubsection{Limitations of and corrections to the Fokker-Plank equation}
In the considered Focker Planck equation, $ P_{\uparrow \uparrow}^{(\bf u)}$ and $ P_{\downarrow \downarrow}^{(\bf u)}$ are decoupled and thus they both preserve their weight ($\sum_{m,m'}  P_{\uparrow \uparrow}^{(\bf u)}(t)=\sum_{m,m'} P_{\downarrow \downarrow}^{(\bf u)} (t)=\frac{1}{2})$. However, numerical results using the exact equation (\ref{secondeqmotiontwoobservables}) (see for example Figs. \ref{Fi:plotPIIdfas} and \ref{Fi:plotPIIdfasII} ), show that the narrow peak at $m=m'=0$, identified with $P_{\downarrow \downarrow}^{(\bf u)}$, decreases until it disappears. The reason for such an effect turns out to be the corrections in (\ref{FPequationTwoApparatusess}) of order $\mathcal{O}(1/m^2N^2,1/m'^2N'^2)$. These terms can be neglected for the initial conditions, in which $m,m'\sim \mathcal{0}(1/\sqrt{N},\sqrt{N'})$, and for large $m$. Thus it holds for the evolution of $ P_{\uparrow \uparrow}^{(\bf u)}$.  However, $P_{\downarrow \downarrow}^{(\bf u)}$ tends to concentrate around $m=m'=0$, reaching an stable given by (\ref{equilibriumdistributionP-}), in which $m,m'\sim \mathcal{O}(1/N,1/N')$. Then, the corrections of order $\mathcal{O}(1/m^2N^2,1/m'^2N'^2)$ become relevant and they couple $ P_{\uparrow \uparrow}^{(\bf u)}$ and $P_{\downarrow \downarrow}^{(\bf u)}$.  

If we account for the corrections, which are given explicitly in (\ref{derivationFPEq}), the Fokker-Plank equation becomes:

\begin{widetext}
\begin{align}
   \frac{\partial P_{ii}^{(\bf u)}}{\partial t}=& \left(\frac{\partial}{\partial m}[-v_{i} P_{ii}^{(\bf u)}]
   +\frac{1}{N}\frac{\partial^2}{\partial m^2}[w_{i} P_{ii}^{(\bf u)}] \right) \left( 1 \mp \frac{2u_x^2u_z^2}{m^2N^2} \right) 
  + \frac{2u_x^2u_z^2}{m^2N^2} \left(w_{j}P_{jj}-w_{i}P_{ii}  \pm \frac{\partial}{\partial m}[-v_{j} P_{jj}] \pm \frac{1}{N}\frac{\partial^2}{\partial m^2}[w_{j} P_{jj}] \right)  
  \nonumber\\
 &+ \mathrm{B'\hspace{1mm} terms} ,
\end{align}
\end{widetext}
with $i,j= \{ \uparrow, \downarrow \}$. 
From these corrections, the term $\frac{2u_x^2u_z^2}{m^2N^2} \left(w_{j}P_{jj}-w_{i}P_{ii} \right)$ is particularly important, because it shows how 
$P_{\downarrow \downarrow}^{(\bf u)}$ is transfered to $ P_{\uparrow \uparrow}^{(\bf u)}$ and vice versa. 
In the case of two identical apparatuses with the simplification (\ref{simplficationw}), we have that 
\begin{align}
 &\frac{2u_x^2u_z^2}{m^2N^2} \left(w_{\mp}P_{\mp}-w_{\pm}P_{\pm} \right)+\frac{2u_x^2u_z^2}{m^{'2}N^2} \left(w^{'}_{\mp}P_{\mp}-w^{'}_{\pm}P_{\pm} \right)\nn\\
 &=\frac{2w}{r^2 N^2}(P_{\mp}-P_{\pm}) ,
 \nonumber\\
& \frac{2u_x^2u_z^2}{m^2N^2} \left(w_{j}P_{jj}^{(\bf u)}-w_{i}P_{ii}^{(\bf u)} \right)+\frac{2u_x^2u_z^2}{m^{'2}N^2} \left(w^{'}_{j}P_{jj}^{(\bf u)}-w^{'}_{i}P_{ii}^{(\bf u)} \right)
\nn\\
& =\frac{2w}{r^2 N^2}(P_{jj}^{(\bf u)}-P_{ii}^{(\bf u)}) ,
\end{align}
with $i,j= \{ \uparrow, \downarrow \}$. Then multiplying and integrating over $r$ the equation of motion:
\begin{equation}
  \frac{\partial}{\partial t}\int_0^{\infty} r  P_{ii}^{(\bf u)} \sim \frac{2w}{ N^2} \int_0^{\infty} \frac{1}{r}(P_{jj}^{(\bf u)}-P_{ii}^{(\bf u)}) ,
\label{transferP-P+}
\end{equation}
with $i,j= \{ \uparrow, \downarrow \}$. 
 In the right hand side the term that contains $P_{\downarrow \downarrow}^{(\bf u)}$ is much bigger than the one with $ P_{\uparrow \uparrow}^{(\bf u)}$, because $P_{\downarrow \downarrow}^{(\bf u)}$ is centered at $r=0$ whereas $ P_{\uparrow \uparrow}^{(\bf u)}$ moves away from the center. Therefore, the transfer between $ P_{\uparrow \uparrow}^{(\bf u)}$ and $P_{\downarrow \downarrow}^{(\bf u)}$ goes basically in one way: from $P_{\downarrow \downarrow}^{(\bf u)}$ to $ P_{\uparrow \uparrow}^{(\bf u)}$. This allows us to describe the disappearance of the peak at $(m=0,m'=0)$ which was observed in the numerical simulations.

The velocity with which $P_{\downarrow \downarrow}^{(\bf u)}$ is transferred to $ P_{\uparrow \uparrow}^{(\bf u)}$ looks very similar to the expression obtained for the velocity with which $P_{\downarrow \downarrow}^{(\bf u)}$ loses memory of its initial conditions, found in (\ref{ratelostmemory}). Indeed, they are just the same except for a factor $1/N$. Therefore, for sufficiently long $N$, we see that first $P_{\downarrow \downarrow}^{(\bf u)}$ becomes symmetric and then it is transferred to $ P_{\uparrow \uparrow}^{(\bf u)}$.

\subsection{Summary of the dynamics}

In conclusion, we have derived a Fokker-Plank equation for $P(m,m',t)$, the probability for the magnets to have magnetizations $m$ and $m'$, respectively, which  allows to describe the time evolution.  The Fokker-Plank equation is characterized by a (two dimensional) field velocity $(v,v')$ and a dispersion $(w,w')$. The dynamics of the probability distribution $P(m,m',t)$ has the following main features:
\begin{itemize}
 \item $P$, whose initial distribution is a gaussian centered at $(0,0)$, splits into two distributions: $ P_{\uparrow \uparrow}^{(\bf u)}$ and $P_{\downarrow \downarrow}^{(\bf u)}$. 
 \item The field velocity of the Fokker-Plank equation shows how $P_{\downarrow \downarrow}^{(\bf u)}$ tends to the center whereas $ P_{\uparrow \uparrow}^{(\bf u)}$ tends to the corners $(\pm m_F,\pm m_F)$. These considerations are in perfect agreement with the ones regarding the free energies $F_{\uparrow \uparrow}^{(\bf u)}$, $F_{\downarrow \downarrow}^{(\bf u)}$. 
\item The dispersion $(w,w')$ tends to symmetrize the distributions $ P_{\uparrow \uparrow}^{(\bf u)}$ and $P_{\downarrow \downarrow}^{(\bf u)}$ (they lose their angular dependence). This has been shown for a simplified scenario, where $(v,v')$ is radial and $w=w'$ is a constant, leading to the result (\ref{ratelostmemory}). Such a symmetrization is much stronger for $P_{\downarrow \downarrow}^{(\bf u)}$ than for $ P_{\uparrow \uparrow}^{(\bf u)}$. 
\item $P_{\downarrow \downarrow}^{(\bf u)}$ is transferred to $ P_{\uparrow \uparrow}^{(\bf u)}$ and viceversa, which is quantified in (\ref{transferP-P+}). This transfer happens mainly in one direction, namely from $P_{\downarrow \downarrow}^{(\bf u)}$ to $ P_{\uparrow \uparrow}^{(\bf u)}$, so that at the end of the process $P_{\downarrow \downarrow}^{(\bf u)}$ has disappeared. The time scale where this process takes place is of the order of $1/N$ of the time scale of the process of symmetrization.
\item Since at the end of the process $P_{\downarrow \downarrow}^{(\bf u)}$ has disappeared, all the probability distribution $P$ is peaked in the corners. The weight of such peaks is an interplay between the field velocity, which tens to send the $ P_{\uparrow \uparrow}^{(\bf u)}$ towards one of the corners (without losing its angular dependence), and the dispersion, which tends to symmetrize the distribution. 

\end{itemize}

\subsection{Form of the state in time}
\label{AppendixForm}

Here we justify expression \eqref{DMM}, i.e., that the state remains in time a function of $\hat{m}$ and $\hat{m}'$. The idea behind is to note that (i) the initial state is a symmetric function of the operators $\sigma_z^{(n)}$ and $\sigma_z^{(n)'}$, and can be written as a function of $\hat{m}$, $\hat{m}'$, and that (ii) this symmetry is not broken, neither by the Hamiltonian of M and M', nor by the coupling MB and M'B', as the baths couples homogeneously to all degrees of freedom of M and M'. Let us here provide a more explicit justification.

 First, let us note that the analogous version for one apparatus, equation \eqref{DMM}, has been justified in the Appendix B of \cite{OpusABN}. Here we use a similar reasoning. Let us take as a starting point the equation of motion \eqref{expbathsi}, given by,
 \begin{align}
 i\hbar \frac{d \hat{D}}{dt} =& [\hat{H}_{SM}+\hat{H}_{SM'},\hat{D}]+\tr_{\rm B} [\hat{H}_{MB},\tr_{B'} \hat{\mathcal{D}}] \nn\\
 +&\tr_{B'} [\hat{H}_{MB'},\tr_{B} \hat{\mathcal{D}}]
\label{2formalexpII}
\end{align}
 which holds for arbitrary initial states. In it we find terms of the form,
 \begin{eqnarray}
\hspace{-7mm} \tr_{\rm B} [\hat{H}_{MB},\tr_{B'} \hat{\mathcal{D}}]&=&\sum_n \int_0^t du\Big\{\Big(\Big[\hat{\sigma}_{+}^{(n)}(u)\hat{D},\hat{\sigma}_-^{(n)}\Big] \\
&+&\Big[\hat{\sigma}_-^{(n)}(u)\hat{D},\hat{\sigma}_+^{(n)}\Big]\Big)K(u)+h.c.\Big\}, \nn
\label{expbathsiII}
\end{eqnarray}
with,
\begin{eqnarray}
\hspace{-9mm}\hat{\sigma}_+^{(n)}(t)= \hat{\sigma}_+^{(n)} g(\hat{m},t)
\label{expevsigmaunII}
\end{eqnarray}
where 
\begin{align}
g(\hat{m},t)=\exp \hspace{-0.6mm} \left\{i\frac{2t}{\hbar}\left(J\hat{m}^3+g u_z {\bf u \cdot \hat{s}}\right)\hspace*{-0.5mm}\right\}.
\end{align}
Similarly, we can find $\hat{\sigma}_-^{(n)}(t)$ by using $\sigma_-=\sigma_+^{\dagger}$. With identical arguments we can compute $\hat{\sigma}_+^{(n)'}(t)$, $\hat{\sigma}_-^{(n)'}(t)$, arising from the second magnet. 
Now, note  that,
\begin{align}
\sigma_+^{(n)} \hat{m} =  (\hat{m} - \frac{2}{N} \mathbb{I}) \sigma_+^{(n)}
\end{align} 
from which if follows,
\begin{align}
\sigma^{(n)}_{\pm} f(\hat{m}, \hat{m}') = f(\hat{m}\mp \delta m, \hat{m}')
\end{align}
where $\delta m=2/N$, and similarly for $\hat{m}'$.   Using this property and \eqref{expevsigmaunII}, and assuming that $\hat{D}=\hat{D}(\hat{m},\hat{m}')$, we notice that  $\hat{\sigma}_+^{(n)'}(t)$, $\hat{\sigma}_-^{(n)'}(t)$ only enter into the right hand side of \eqref{2formalexpII} through combinations of the form $\hat{\sigma}_+^{(n)}\hat{\sigma}_-^{(n)}$. Summing over all $n$ as in \eqref{expbathsiII}, we find,
\begin{align}
\sum_n \sigma_+^{(n)} \sigma_-^{(n)} = \frac{N}{2} (\mathbb{I}) +\hat{m}).
\end{align}
This shows that the right hand side of \eqref{2formalexpII} only depends on $\hat{m}$ if $\hat{D}=\hat{D}(\hat{m},\hat{m}')$. Identical arguments lead to the conclusion that it also depends only on $\hat{m}'$. Hence we conclude that the evolution of $\hat{D}$ can be expressed as $\hat{D}(\hat{m},\hat{m}',t)$ as long as the initial state can be expressed as a function of $\hat{m}$ and $\hat{m}'$.

 \renewcommand{\thesection}{\Alph{section}}
  \renewcommand{\thesubsection}{\thesection\arabic{subsection}}
\section{Determining the minimal coupling for both apparatuses to register results}
 \renewcommand{\thesection}{\Alph{section}.}

\label{AppMinimalCoupling}

In this section we aim to find the minimal coupling $h_d$ for which $F_{\Uparrow}^{(\bf u)}$ contains no free energy barriers.  This is equivalent to demanding, 
\begin{align}
& \frac{\partial F_{\Uparrow}^{(\bf u)}}{\partial m} \leq 0\hspace{5mm} \forall m\in (0,m_F) , \nonumber\\ 
& \frac{\partial F_{\Uparrow}^{(\bf u)}}{\partial m'} \leq 0 \hspace{5mm} \forall m' \in (0,m_F) , \nonumber\\
& \frac{\partial F_{\Uparrow}^{(\bf u)}}{\partial m} \geq 0\hspace{5mm} \forall m\in (0,-m_F) , \nonumber\\ 
& \frac{\partial F_{\Uparrow}^{(\bf u)}}{\partial m'} \geq 0 \hspace{5mm} \forall m' \in (0,-m_F) ,
\end{align}
where $m_F \approx 1$, is the value where the ferromagnetic distribution peaks. For the first condition we obtain, 
\begin{align}
\frac{1}{N}  \frac{\partial F_{\Uparrow}^{(\bf u)}}{\partial m}=& - \left(\frac{Ng^2 m}{2\sqrt{(Ngm)^2+(N'g'm')^{2}}}+
    J_2  m+J_4  m^3\right) \nn\\&
+\frac{1}{2\beta} \ln \left(\frac{m+1}{1-m} \right) .
\label{partialF}
\end{align}
This function is an odd function of $m$ -- and similarly for $ \partial F_{\Uparrow}^{(\bf u)} /\partial m'$. Hence the previous conditions reduce to,
\begin{align}
& \frac{\partial F_{\Uparrow}^{(\bf u)}}{\partial m} \leq 0\hspace{5mm} \forall m\in (0,m_F) , \nonumber\\ 
& \frac{\partial F_{\Uparrow}^{(\bf u)}}{\partial m'} \leq 0 \hspace{5mm} \forall m' \in (0,m_F) .
\end{align} 
Going back to \eqref{partialF}, negativity of this function as a function of $g$ becomes most demanding for $m'=1$.  Therefore, to satisfy the previous conditions  it is enough to demand,
\begin{align}
&A(m)\equiv \frac{1}{N} \frac{\partial F_{\Uparrow}^{(\bf u)}}{\partial m} \Bigr|_{m'=1}<0, \hspace{5mm} \forall m\in (0,m_F) , \nonumber\\
&B(m)\equiv \frac{1}{N'} \frac{\partial F_{\Uparrow}^{(\bf u)}}{\partial m'} \Bigr|_{m=1}<0, \hspace{5mm} \forall m'\in (0,m_F).
\end{align}
In the interval $m\in (0,m_F)$, the function $A(m)$ can only become positive if $T\geq J_2$ and for small values of $m$. Negativity of the function can then be ensured by imposing $\partial A(m)/ \partial m |_{m=0} <0$ which leads to the following simple solution for $g'=g$, 
\begin{equation}
h_d=2  \max (T-J_2,T-J_2' ). 
\label{appminimumcouplingfeapp}
\end{equation}


\begin{thebibliography}{0}



\bibitem{other2}
W. H. Zurek, \emph{Decoherence, einselection, and the quantum origins of the classical}, 
\href{http://journals.aps.org/rmp/abstract/10.1103/RevModPhys.75.715}{Rev. Mod. Phys. {\bf 75}, 715 (2003)}.


\bibitem{other3}
M. Schlosshauer,  \emph{Decoherence, the measurement problem, and interpretations of quantum mechanics}, 
\href{http://journals.aps.org/rmp/abstract/10.1103/RevModPhys.76.1267}{Rev. Mod. Phys. {\bf 76}, 1267 (2004)}.

\bibitem{other1} 
F. Lalo\"e, {\it  Comprenons-nous vraiment la m\'ecanique quantique},  EDP Sciences/CNRS Editions, Paris 2011.

\bibitem{OpusABN} A. E. Allahverdyan, R. Balian and Th. M. Nieuwenhuizen, \emph{Understanding quantum measurement from the solution of dynamical models}, 
\href{http://www.sciencedirect.com/science/article/pii/S0370157312004085}{ Phys. Rep. {\bf 525}, 1 (2013)}.

\bibitem{OpusII} A. E. Allahverdyan, R. Balian, T. M. Nieuwenhuizen, \emph{A sub-ensemble theory of ideal quantum measurement processes}, \href{http://dx.doi.org/10.1016/j.aop.2016.11.001}{Annals of Physics, in Press}. 



\bibitem{Shay} S. Hacohen-Gourgy, L. S. Martin, E. Flurin, V. V. Ramasesh, K. B. Whaley, I. Siddiqi, \emph{Dynamics of simultaneously measured non-commuting observables},  
\href{http://www.nature.com/nature/journal/v538/n7626/full/nature19762.html}{Nature {\bf 538}, 491-494 (2016)}.

\bibitem{gills} G. P\" utz, T. Barnea, N. Gisin, and A. Martin, \emph{Experimental weak measurement of two non-commuting observables}, 	. 
\href{https://arxiv.org/abs/1610.04464}{arXiv:1610.04464  (2016)}.

 









 


\bibitem{Arthurs} E. Arthurs and J. l. Kelly, \emph{On the Simultaneous Measurement of a Pair of Conjugate Observables}, Bell Syst. Tech. \textbf{44}, 725 (1965).

\bibitem{Buschh} P. Busch,  \emph{Some realizable joint measurements of complementary observables},  \href{http://dx.doi.org/10.1007/BF00734320}{P. Found Phys 17: 905 (1987). }

\bibitem{BuschLahti} 
 P. Busch, M. Grabowski, and P. J. Lahti, \emph{Operational Quantum Physics},  Springer Lecture Notes in Physics Monographs
Volume 31 (1995).

\bibitem{MuynkI} W. M. de Muynck, \emph{Foundations of quantum mechanics, an empiricist approach}, (Kluwer Academic Publishers, Dordrecht, 2002).

\bibitem{MuynkII} Hans Martens, Willem M. de Muynck, \emph{Nonideal quantum measurements},  \href{https://link.springer.com/article/10.1007/BF00731693 }{Found. of Phys. 20 (1990), p. 255-281}.


\bibitem{Armen} A. E. Allahverdyan, R. Balian and Th. M. Nieuwenhuizen,  \emph{Simultaneous measurement of non-commuting observables}, 
 \href{http://www.sciencedirect.com/science/article/pii/S1386947709003075}{ Physica E Low-dimensional Systems and Nanostructures 60(3) (2010)}.


\bibitem{Ziman}  T. Heinosaari, T. Miyadera, M. Ziman, \emph{An Invitation to Quantum Incompatibility},  %
 \href{http://iopscience.iop.org/article/10.1088/1751-8113/49/12/123001/meta}{J. Phys. A: Math. Theor. {\bf 49}  123001 (2016)}.

 \bibitem{ArmenII}  A.E. Allahverdyan, R. Balian, T.M. Nieuwenhuizen, \emph{Determining a Quantum State by Means of a Single Apparatus}, \href{http://journals.aps.org/prl/abstract/10.1103/PhysRevLett.92.120402}{Phys. Rev. Lett. \textbf{92} (12) 120402  (2004).}

\bibitem{Ozawa} M. Ozawa, \emph{Universally valid reformulation of the Heisenberg uncertainty principle on noise and disturbance in measurement},
 \href{http://journals.aps.org/pra/abstract/10.1103/PhysRevA.67.042105}{ Phys. Rev. A \textbf{67}, 042105 (2003)}.
 

\bibitem{Mohseni}  M. Mohseni, A. M. Steinberg, and J. A. Bergou, \emph{Optical Realization of Optimal Unambiguous Discrimination for Pure and Mixed Quantum States},
\href{10.1103/PhysRevLett.93.200403}{Phys. Rev. Lett. 93, 200403 (2004).}

\bibitem{Andersson} E. Andersson, S. M. Barnett, A. Aspect, \emph{Joint measurements of spin, operational locality, and uncertainty},  
 \href{http://journals.aps.org/pra/abstract/10.1103/PhysRevA.72.042104}{ Phys. Rev. A \textbf{72}, 042104 (2005)}.


\bibitem{Busch}  P. Busch, P. Lahti, R. F.  Werner, \emph{Proof of Heisenberg's error-disturbance relation}, %
 \href{http://journals.aps.org/prl/abstract/10.1103/PhysRevLett.111.160405}{Phys. Rev. Let. \textbf{111} (16), 160405 (2013)}.


\bibitem{BuschReview} P. Busch, P. Lahti, and R. F. Werner, \emph{Colloquium: Quantum root-mean-square error and measurement uncertainty relations},
 \href{http://journals.aps.org/rmp/abstract/10.1103/RevModPhys.86.1261}{Rev. Mod. Phys. \textbf{86}, 1261 (2014)}.


\bibitem{Branciard} C. Branciard, \emph{Deriving tight error-trade-off relations for approximate joint measurements of incompatible quantum observables}, 
\href{http://journals.aps.org/pra/abstract/10.1103/PhysRevA.89.022124}{Phys. Rev. A \textbf{89}, 022124, (2014)}.

\bibitem{Bell} N. Brunner, D. Cavalcanti, S. Pironio, V. Scarani, and S. Wehner, \emph{Bell nonlocality}, 
\href{http://journals.aps.org/rmp/abstract/10.1103/RevModPhys.86.419}{Reviews of Modern Physics {\bf 86}, 419 (2014)}.


\bibitem{Wolf} M. M. Wolf, D. Perez-Garcia, and C. Fernandez, \emph{Measurements Incompatible in Quantum Theory Cannot Be Measured Jointly in Any Other No-Signaling Theory}, 
\href{http://journals.aps.org/prl/abstract/10.1103/PhysRevLett.103.230402}{Phys. Rev. Lett. 103, 230402 (2009)}.

\bibitem{Marco} M. T. Quintino, T. V\' ertesi, N. Brunner \emph{Joint Measurability, Einstein-Podolsky-Rosen Steering, and Bell Nonlocality}, 
\href{http://journals.aps.org/prl/abstract/10.1103/PhysRevLett.113.160402}{Phys. Rev. Lett. 113, 160402 (2014)}.

\bibitem{Uola} R. Uola, T. Moroder, and O. G\" uhne,  \emph{Joint Measurability of Generalized Measurements Implies Classicality},  
\href{http://journals.aps.org/prl/abstract/10.1103/PhysRevLett.113.160403}{Phys. Rev. Lett. 113, 160403 (2014)}.


\bibitem{BuschQubit} P. Busch, P. Lahti,  and R. F. Werner, \emph{Heisenberg uncertainty for qubit measurements}, \href{http://link.aps.org/doi/10.1103/PhysRevA.89.012129}{Phys. Rev. A. 89, 012129 (2014)}.

\bibitem{BranciardQubit} A. Abbott, and C. Branciard, \emph{Noise and disturbance of qubit measurements: An information-theoretic characterization}, \href{http://link.aps.org/doi/10.1103/PhysRevA.94.062110}{Phys. Rev. A {\bf 13} 062110 (2016)}.


\bibitem{ei}  M. Kac, \emph{Mathematical Mechanisms of Phase Transitions} in: Statistical Mechanics of Phase Transitions and Superfluidity, Gordon and Breach Science Publishers, New York, 1968. 

\bibitem{eii}  A. Bovier and I. Kurkova, \emph{A Short Course on Mean Field spin glasses} in: Spin Glasses: Statics and Dynamics, Birkh\"auser Basel, (2009), p.3. 

\bibitem{eiii} M. Kochma\'nski, T. Paszkiewicz, and S. Wolski, \emph{Curie-€"Weiss magnet€"-a simple model of phase transition}, \href{http://stacks.iop.org/0143-0807/34/i=6/a=1555}{European Journal of Physics, 34,  6, 1555(2013)}.

 \bibitem{ABNCW} A. E. Allahverdyan, R. Balian and Th. M. Nieuwenhuizen,  \emph{Curie-Weiss model of the quantum measurement process}, 
 \href{http://iopscience.iop.org/article/10.1209/epl/i2003-00150-y/meta}{ Europhys. Lett. \textbf{61}, 453 (2003)}.

\bibitem{Ruskov} R. Ruskov, A. N. Korotkov, and K. M\o lmer, \emph{Qubit State Monitoring by Measurement of Three Complementary Observables}, 
\href{http://journals.aps.org/prl/abstract/10.1103/PhysRevLett.105.100506}{Phys. Rev. Lett. 105, 100506 (2010)}.

\bibitem{Jordan} A. N. Jordan and  M. B\"uttiker, \emph{Continuous quantum measurement with independent detector cross correlations},  
\href{http://journals.aps.org/prl/abstract/10.1103/PhysRevLett.95.220401}{Phys. Rev. Lett. 95, 220401 (2005)}.

%


\bibitem{Luis} L. P.  Garc\'ia-Pintos, J. Dressel, \emph{Probing quantumness with joint continuous measurements of non-commuting qubit observables}, 	
\href{https://arxiv.org/abs/1606.07934}{arXiv:1606.07934 (2016)}.

\bibitem{Mitchison} G. Mitchison, R. Jozsa, and S. Popescu, \emph{Sequential weak measurement},  
\href{http://journals.aps.org/pra/abstract/10.1103/PhysRevA.76.062105}{Phys. Rev. A {\bf 76}, 062105 (2007)}.










 \bibitem{breuer} HP Breuer, F Petruccione, \emph{The theory of open quantum systems},  (Oxford
University Press, Oxford, 2002).(2002).

\bibitem{Cirac} E. M. Kessler,  G. Giedke, A. Imamoglu, S. F. Yelin, M.D.  Lukin, and J. I. Cirac, \emph{Dissipative phase transition in a central spin system}, \href{http://link.aps.org/doi/10.1103/PhysRevA.86.012116}{Phys. Rev. A {\bf 86} 012116 (2012)}




 
 \bibitem{Nielsen} M. A. Nielsen, I. L. Chuang, \emph{Quantum computation and quantum information},  Cambridge University Press (2000).


\bibitem{marti} M. Perarnau-Llobet, \emph{Master Thesis}, University of Amsterdam (2012).

 
 
 
\bibitem{commentR} Equality is only strictly true in the limit $N\rightarrow \infty$. Otherwise, $R_{\Uparrow}$ and $R_{\Downarrow}$ correspond to gaussian states with average $\pm m_F$ magnetization. 
 

\bibitem{footnoteg}  Here we are also implicitly assuming that the temperature is low enough, $T<0.363J$, so that the paramagnetic state is metastable. 


\end{thebibliography}
\end{document}